\documentclass[notitlepage, 11pt]{article}

\usepackage{color}
\usepackage{amsmath}
\usepackage{latexsym,amssymb}
\usepackage{graphicx}

\newtheorem{theorem}{Theorem}

\newtheorem{example}{Example}
\newtheorem{proposition}{Proposition}

\newtheorem{remark}{Remark}

\newcommand {\reals} {{\rm I\!R}}

\def\thelemma{\arabic{section}.\arabic{lemma}}
\def\thetheorem{\arabic{section}.\arabic{theorem}}
\def\thecorollary{\arabic{section}.\arabic{corollary}}
\def\thedefinition{\arabic{section}.\arabic{definition}}
\def\theexample{\arabic{section}.\arabic{example}}
\def\theproposition{\arabic{section}.\arabic{proposition}}

\def\theassumption{\arabic{section}.\arabic{assumption}}

\def\theremark{\arabic{section}.\arabic{remark}}

\newcommand{\manualnames}[1]{\def\theequation{#1.\arabic{equation}}
\def\thelemma{#1.\arabic{lemma}}
\def\thetheorem{#1.\arabic{theorem}}
\def\thecorollary{#1.\arabic{corollary}}
\def\thedefinition{#1.\arabic{definition}}
\def\theexample{#1.\arabic{example}}
\def\theproposition{#1.\arabic{proposition}}
\def\theassumption{#1.\arabic{assumption}}
\def\theremark{#1.\arabic{remark}}
}

\newcommand{\beginsec}{
\setcounter{lemma}{0}
\setcounter{theorem}{0}
\setcounter{corollary}{0}
\setcounter{definition}{0}
\setcounter{example}{0}
\setcounter{proposition}{0}
\setcounter{condition}{0}
\setcounter{assumption}{0}
\setcounter{conjecture}{0}
\setcounter{problem}{0}
\setcounter{remark}{0}
}

\newcommand{\dfn}{\stackrel{\triangle}{=}}

\newcommand{\al}{\alpha}

\newcommand{\s}{\sigma}
\newcommand{\del}{\delta}
\newcommand{\om}{\omega}

\newcommand{\Gam}{\mathnormal{\Gamma}}
\newcommand{\Del}{\mathnormal{\Delta}}

\newcommand{\PI}{\mathnormal{\Pi}}
\newcommand{\Sig}{\mathnormal{\Sigma}}
\newcommand{\Ph}{\mathnormal{\Phi}}

\newcommand{\Om}{\mathnormal{\Omega}}

\newcommand{\C}{{\mathbb C}}

\newcommand{\R}{\mathbb{R}}
\newcommand{\N}{\mathbb{N}}

\newcommand{\la}{\lambda}
\newcommand{\sig}{\sigma}
\newcommand{\eps}{\epsilon}

\newcommand{\calA}{{\cal A}}

\newcommand{\calC}{{\cal C}}

\newcommand{\calE}{{\cal E}}
\newcommand{\calF}{{\cal F}}

\newcommand{\calL}{{\cal L}}
\newcommand{\calM}{{\cal M}}
\newcommand{\calN}{{\cal N}}
\newcommand{\calP}{{\cal P}}

\newcommand{\calS}{{\cal S}}
\newcommand{\calT}{{\cal T}}

\newcommand{\calX}{{\cal X}}
\newcommand{\calY}{{\cal Y}}

\DeclareMathOperator*{\esssup}{ess\,sup}

\newcommand{\oo}{\overline}
\newcommand{\skp}{\vspace{\baselineskip}}

\newcommand{\iy}{\infty}

\newcommand{\ds}{\displaystyle}

\newcommand{\proof}{\noindent{\bf Proof:}\ }
\newcommand{\qed}{\hfill $\Box$}

\newcommand {\ba} {\mbox{\boldmath $a$}}

\newcommand {\bc} {\mbox{\boldmath $c$}}

\newcommand {\bh} {\mbox{\boldmath $h$}}

\newcommand {\bx} {\mbox{\boldmath $x$}}
\newcommand {\by} {\mbox{\boldmath $y$}}
\newcommand {\bz} {\mbox{\boldmath $z$}}
\newcommand {\bA} {\mbox{\boldmath $A$}}

\newcommand {\bC} {\mbox{\boldmath $C$}}

\newcommand {\bE} {\mbox{\boldmath $E$}}

\newcommand {\bN} {\mbox{\boldmath $N$}}

\newcommand {\bX} {\mbox{\boldmath $X$}}
\newcommand {\bY} {\mbox{\boldmath $Y$}}
\newcommand {\bZ} {\mbox{\boldmath $Z$}}
\newcommand{\btheta}{\mbox{\boldmath $\theta$}}
\newcommand{\aaa}{{A}}
\newcommand{\dd}{{\rm d}}
\newcommand{\sigmax}{{\Sig_{\rm max}}}

\begin{document}

\title{Information-Theoretic Applications of the
Logarithmic Probability Comparison Bound}
\author{Rami Atar\thanks{Research
supported in part by the ISF (Grant 1315/12).}
\hspace{4em}
Neri Merhav\thanks{Research
supported in part by the ISF (Grant 412/12).}
\\ \\
Department of Electrical Engineering \\
Technion -- Israel Institute of Technology \\
Haifa 3200003, ISRAEL\\
\{atar, merhav\}@ee.technion.ac.il}

\date{December 21, 2014}

\maketitle

\begin{abstract}
%\setlength{\baselineskip}{0.5\baselineskip}

%The logarithmic probability comparison bound (LPCB), which
%is based on a change--of--measure inequality, has recently been proposed by
%Atar, Chowdhary, and Dupuis \cite{ACD14}
%with various applications in probability theory
%and stochastic processes. When used to bound
%exponential rates of probabilities of rare events, it
%can be viewed as an extension of the well known
%idea (used, e.g., in the sphere--packing bound) of passing from the original
%probability measure to a new measure under which the probability of the
%event in question becomes non--exponential. The purpose of this paper is to
%present the LPCB and to demonstrate its usefulness in various
%information--theoretic applications.

A well-known technique in estimating probabilities of rare events in general and in
information theory in particular (used, e.g., in the sphere--packing bound),
is that of finding a reference probability measure
under which the event of interest has probability of order one
and estimating the probability in question
by means of the Kullback-Leibler divergence. A method has recently been
proposed in \cite{ACD14}, that can be viewed as an extension
of this idea in which the probability under the reference measure
may itself be decaying exponentially, and the R\'enyi divergence is used instead.
The purpose of this paper is to demonstrate the usefulness of this
approach in various information--theoretic settings.
For the problem of channel coding, we provide a general methodology for
obtaining matched, mismatched and robust error exponent bounds,
as well as new results in a variety of particular channel models.
Other applications we address include rate-distortion coding and the problem of guessing.

\vspace{0.2cm}


\end{abstract}

%\setlength{\baselineskip}{2\baselineskip}
%\newpage

\section{Introduction}

A key approach to obtaining lower bounds on probabilities of rare
events is based on the idea of a change of measure. In this approach,
the underlying
probability measure is replaced by a reference probability measure under
which the probability of the event in question does not decay exponentially,
and the
exponent of the bound is given by the Kullback--Leibler (KL) divergence between the
two probability measures. One then
optimizes the estimate over all reference measures having the property
alluded to above.
This idea is standard for
deriving lower bounds in large deviations theory
(see, e.g., \cite[p.\ 32]{DZ}), where it is sometimes
referred to as tilting. In the context of information theory it has been used in
applications including (i) the derivation of the sphere--packing bound
for discrete memoryless channels (DMC's), using
Csisz\'ar and K\"orner's method \cite[Theorem 5.3]{CK};
(ii) Marton's converse theorem on the source coding
exponent \cite{Marton74}. In the former, the
resulting exponential error bound is tight at least in some range of high
coding rates. In the latter, it is virtually always tight (for finite--alphabet memoryless
sources), as there exists a matching upper bound.

In \cite{ACD14}, Atar, Chowdhary and Dupuis presented what may be viewed as an
extension of this approach to situations where
the probability of the event of interest may also decay exponentially under the
reference measure.
The estimate is then given in terms of the corresponding R\'enyi divergence.
At the heart of the approach lies the
{\it logarithmic probability comparison bounds} (LPCB)
that compare the probability of an event under two measures at a logarithmic
scale in terms of the respective R\'enyi divergence. Specifically, if $P$ and $Q$
are probability measures on a measurable space and $\calA$ is an event on it then
\begin{equation}\label{76}
  \frac{1}{\al-1}\ln P(\calA)\le\frac{1}{\al}\ln Q(\calA)+D_\al(P\|Q)
\end{equation}
for $\al>1$, where $D_\al$ denotes R\'enyi divergence of order $\al$
(see definition and details in Section \ref{sec2}).
This bound is tight in the sense that, given $P$ and $\calA$, one can find $Q$
for which it holds as equality. Thus if $\{P_n\}$
and $\{Q_n\}$ are sequences of probability measures and we denote by
$E_P=-\limsup_{n\to\iy} n^{-1}\ln P_n(\calA)$ the exponential decay rate of the probability under
$P_n$ and by $E_Q$ that under $Q_n$, then with
$\Del_\al=\limsup_{n\to\iy} n^{-1} D_\al(P_n\|Q_n)$, one obtains
\begin{equation}
  \label{77}
  E_P\ge \frac{\al-1}{\al}E_Q-(\al-1)\Del_\al,\qquad \al>1.
\end{equation}
This gives a lower bound on the decay rate $E_P$ under a sequence of measures of interest
in terms of that under reference measures, $E_Q$.
By switching the roles of $\{P_n\}$ and $\{Q_n\}$ one obtains an analogous upper bound.
One natural use of \eqref{77} is when $Q$ is a model for which
we have information on the decay rate (exactly or as a bound),
whereas $P$ is harder to analyze.
In this case, a key step is to provide a useful estimate of the divergence term $\Del_\al$.
Another way to view \eqref{77} is as what is often called a {\it robust
bound}, where one attempts to obtain performance bounds on a whole
family of true models $P$, and $Q$ serves in defining this family.
For example, the family of true models might consist of all $P$ for which
the divergence from $Q$ does not exceed a certain bound,
in the sense that $\Del_\al\le \eps(\al)$, some $\eps(\cdot)$.
Then it is immediate from
\eqref{77} that for all $P$ in the family,
\begin{equation}
E_P\ge \frac{\al-1}{\al}E_Q-(\al-1)\eps(\al).
\end{equation}
While the latter has been the main motivation in \cite{ACD14},
both viewpoints will be addressed in this paper.
Some benefits of the approach include: (i) the ability to compare, not only
probabilities of a given event, but also expectations of a given function
under the two (sequences of) measures
(this relies on a more general inequality than \eqref{76}; see
Section \ref{sec2}), (ii) the presence of the free
parameter $\al$, that can be optimized over in order to tighten
the bound, and (iii) the possibility to derive both upper and lower bounds
by the same method.

The objective of this paper is to present the LPCB and the aforementioned method
to the Information Theory audience and to demonstrate its
power and usefulness as a tool for deriving upper and lower bounds
in a variety of applications, including both source
coding and channel coding scenarios. Because it compares two probability measures, the
bound is especially natural to apply in situations of mismatch between the
true underlying model and the one to which the coding--decoding schemes are tailored.
Also, as will be seen in the sequel, in most of these
applications, the setting is sufficiently general that no alternative
bounds are available to the best knowledge of the authors, such as, for example,
coding for channels with additive interference of unlimited memory and mismatch.
In some of these
scenarios, the exponential bounds obtained are tight in the sense that they are
attained at least for some instance of the problem.

Our main contributions are summarized as follows.
\begin{itemize}
\item Highlighting the relevance of the approach to information theory;
\item Developing general
   upper and lower bounds on channel coding error exponents for the
   matched, mismatched and robust settings based on the LPCB;
\item Using the approach to derive new bounds on error exponents
   for a host of particular channel models including
   Gaussian channels with long memory interference,
   the inter-symbol interference channel,
   the fading channel, and the binary erasure channel;
\item Obtaining new bounds for source coding and the problem of guessing.
\end{itemize}

The outline of the remaining part of this paper is as follows. In Section 2,
we present the LPCB. In Section 3, we explain its use in estimating probabilities
of rare events. We also present a corollary regarding small perturbations
between reference and true models.
Section 4 is devoted to the channel coding framework.
Finally, Section 5 provides further application examples.

\skp

{\it Notation.} A vector (deterministic or random) of the form
$(x_1,x_2,\ldots, x_n)$ will be written as $x^n$. When the dimension
$n$ is understood from the context, the vector will sometimes be written as
the corresponding bold font letter, $\bx$.
The probability law of a random variable $X$ under a probability measure $P$ is denoted
by $P_X$, and the conditional law of $Y$ given $X$ under $P$ by
$P_{Y|X}(\cdot|\cdot)$. When there is no room for ambiguity, these
subscripts will be omitted.
Expectation with respect to a probability measure $P$ will be
denoted by $\bE_P\{\cdot\}$. Again, the subscript will be omitted if the underlying
probability distribution is clear from the context.
The entropy of a distribution $Q$ will be denoted by $H(Q)$.

\section{R\'enyi divergence and the LPCB}\label{sec2}

Let a measurable space $(\calS,\calF)$ be given, and denote by $\calP$
the set of all probability measures on it.
For $\al>1$ and $P,Q\in\calP$,
the R\'enyi divergence of degree $\al$ of $Q$ from $P$ is defined by\footnote{Some
authors use the factor $\frac{1}{\al-1}$ rather than $\frac{1}{\al(\al-1)}$.
By choosing the latter we follow the notation used in \cite{lievaj}.}
\begin{equation}
  \label{12}
  D_\al(Q\|P)=\begin{cases}\ds
     \frac{1}{\al(\al-1)}\ln\int\Big(\frac{\dd Q}{\dd P}\Big)^\al \dd P
     & \text{if } Q\ll P,\\
     +\iy & \text{otherwise,}
  \end{cases}
\end{equation}
where $Q\ll P$ denotes absolute continuity of $Q$ with respect
to $P$, and $\frac{\dd Q}{\dd P}$ denotes the Radon-Nikodym derivative.
For $\al=1$ one extends this definition by letting $D_1=D$ be
the KL divergence, namely
\begin{equation}
  \label{13}
  D(Q\|P)=\begin{cases}\ds
     \int\Big(\ln\frac{\dd Q}{\dd P}\Big)\, \dd Q
     & \text{if } Q\ll P,\\
     +\iy & \text{otherwise.}
  \end{cases}
\end{equation}
For $Q$ and $P$ fixed, $\al\mapsto\al D_\al(Q\|P)$
is nondecreasing as a map from $[1,\iy)$ to $[0,\iy]$. Moreover,
if $\bar\al:=\sup\{\al:D_\al(Q\|P)<\iy\}$ and $\bar\al>1$ then
$D_\al(Q\|P)$ is finite and continuous on $[1,\bar\al)$.
For extension to $\al\in\R$ and many other useful properties of the divergence,
see \cite{golpasyar}, \cite{lievaj}, \cite{vaj} and \cite{ervhar}.

The well-known convex duality between exponential integrals
and KL divergence \cite{dupell4} states that for any bounded measurable function
$g:\calS\to\R$, and every $Q\in\calP$,
\begin{equation}
  \label{14}
  \ln\int e^g\dd Q=\sup_{P\in\calP}\Big[\int g\dd P-D(P\|Q)\Big].
\end{equation}

It has recently been shown (in \cite{ACD14}; earlier related calculations
appeared in \cite{dvitod}) that
\begin{equation}
  \label{15}
  \frac{1}{\al}\ln\int e^{\al g}\dd Q=\sup_{P\in\calP}
  \Big[\frac{1}{\al-1}\ln\int e^{(\al-1)g}\dd P-D_\al(P\|Q)\Big], \qquad \al>1.
\end{equation}
Formally, one can recover \eqref{14} from \eqref{15} by taking the limit $\al\to1$ and using
$D_1$ in place of the limit of $D_\al$ as $\al\to1$.
Now, as a consequence of \eqref{15} one obtains for $\al>1$ and $P,Q\in\calP$
the bound
\begin{equation}
  \label{16}
  \frac{1}{\al-1}\ln\int e^{(\al-1)g}\dd P\le
  \frac{1}{\al}\ln\int e^{\al g}\dd Q+D_\al(P\|Q).
\end{equation}
Given an event $\calA\in\calF$, one can take $g$ to assume the values $0$ and $-M$ on
$\calA$ and its complement, respectively, and on taking the limit
$M\to\iy$, deduce from the above that
\begin{equation}
  \label{17}
  \frac{1}{\al-1}\ln P(\calA)\le\frac{1}{\al}\ln Q(\calA)+D_\al(P\|Q)
\end{equation}
(see \cite{ACD14} for the details).
Inequalities \eqref{16} and \eqref{17} are referred to in \cite{ACD14}
as the {\it risk-sensitive functionals comparison bound} and
{\it logarithmic probability comparison bound}, respectively.
It is important to mention that both inequalities are tight in the sense
that given $Q$ and $g$, there exists a (unique) measure, namely $dP=e^{\al g}dQ/Z$,
$Z=\int e^{\al g}\dd Q$, for which \eqref{16} holds with equality.
And given $Q$ [resp., $P$] and $\calA$ for which $Q(\calA)>0$ [resp., $P(\calA)>0$],
there exists a (unique) measure, namely
$Q(\cdot|\calA)$ [resp., $P(\cdot|\calA)$] for which \eqref{17} holds with equality.
Another useful fact is that both also give lower bound in addition to an upper bound,
by interchanging the roles of the measures. Thus
\begin{equation}
  \label{18}
  \frac{1}{\al}\ln P(\calA)\ge\frac{1}{\al-1}\ln Q(\calA)-D_\al(Q\|P).
\end{equation}

\section{Implications on exponential rate of decay}\label{sec3}

It is well known that
\eqref{14} can be used to obtain estimates on probabilities of rare events
(see \cite{dupell4}). By an approach developed in
\cite{ACD14}, the representation
\eqref{15} also leads to such estimates, by appealing to \eqref{16} and \eqref{17}.
We now present this approach. Consider first the simple case where a sequence
of real valued random variables $X_1,X_2,\ldots$ defined on the given measurable space
is i.i.d.\  under both probability measures $P$ and $Q$. Denote by $P_n$ and $Q_n$
the respective probability laws of the vector $X^n=(X_1,\ldots,X_n)$.
It is a simple fact that the R\'enyi divergence scales as
$D_\al(P_n\|Q_n)=nD_\al(P_1\|Q_1)$. Thus for
$n\in\N$ and any event $\calA_n$ measurable on the sigma-field generated
by $(X_1,\ldots, X_n)$, that is, for some Borel subset $B_n$ of $\R^n$,
$\calA_n=\{X^n\in B_n\}$, one has
\begin{equation}
  \limsup_{n\to\iy}
  \frac{1}{n}\ln P(\calA_n)\le\frac{\al}{\al-1}\limsup_{n\to\iy}\frac{1}{n}\ln Q(\calA_n)
  +\al D_\al(P_1\|Q_1).
\end{equation}
This gives a comparison of the exponential rates
involving only the R\'enyi divergence between the two marginals.
In greater generality, when $X_n$ are not necessarily i.i.d.\  under the measures
$P$ and $Q$, with $P_n$ and $Q_n$ still denoting the respective probability
laws of $(X_1,\ldots,X_n)$,
for $n\in\N$, let $G_n$ and $\calA_n$ be a bounded,
measurable function and an event, that are both measurable on the
sigma-field generated by that vector. Then again, from \eqref{16} and \eqref{17},
\begin{equation}
  \label{31}
  \frac{1}{(\al-1)n}\ln \bE_P[e^{(\al-1)G_n(X^n)}]
  \le\frac{1}{\al n} \ln \bE_Q[e^{\al G_n(X^n)}] + \frac{1}{n} D_\al(P_n\|Q_n),
\end{equation}
\begin{equation}
  \label{30}
  \frac{1}{(\al-1)n} \ln P(\calA_n)\le \frac{1}{\al n}\ln Q(\calA_n)
  + \frac{1}{n} D_\al(P_n\|Q_n).
\end{equation}
Denote
\begin{equation}\label{80}
E_*(\hat P)=-\limsup_{n\to\iy}\frac{1}{n}\ln \hat P(\calA_n),
\qquad
E^*(\hat P)=-\liminf_{n\to\iy}\frac{1}{n}\ln \hat P(\calA_n),\qquad \text{for } \hat P\in\calP,
\end{equation}
and
\begin{equation}\label{80+}
\Del_\al^{\hat P,\hat Q}=\limsup_{n\to\iy}\frac{1}{n}D_\al(\hat P_n\|\hat Q_n)
\qquad \text{for } (\hat P,\hat Q)\in\calP^2.
\end{equation}
Then
\begin{equation}
  \label{32}
  \frac{1}{\al-1}E_*(P)\ge\frac{1}{\al}E_*(Q)-\Del_\al^{P,Q}.
\end{equation}
Combining this with the bound obtained by interchanging $P$ and $Q$, one obtains
the two-sided bound on the exponential decay rate under $P$ in terms of that under $Q$:
\begin{equation}\label{52}
  \frac{\al-1}{\al} E_*(Q)-(\al-1)\Del_\al^{P,Q}\le E_*(P)\le E^*(P)
  \le \frac{\al}{\al-1}E^*(Q)+\al\Del_\al^{Q,P}.
\end{equation}
Note that upper and a lower bounds analogous to \eqref{52} can be deduced
from \eqref{31} for limits of the left-hand side of \eqref{31}.
In the sequel, when
the limits exist, we write $E^*(\cdot)$ and $E_*(\cdot)$ as $E(\cdot)$.
We will usually take $P$ to be the model of interest,
or the `true' model, and $Q$ will be the reference model.

It is instructive to note that inequalities \eqref{31} and \eqref{30},
that are valid for each $n$, provide some information that is lost when passing
to the limit, as for example in the i.i.d.\  case alluded to above, where
the divergence term $n^{-1}D_\al(P_n\|Q_n)=D_\al(P_1\|Q_1)$ is given
explicitly. This viewpoint of the approach has been further developed in \cite{ACD14}.
However, in this paper, we will use the bounds exclusively in their limit forms,
given by \eqref{52}.

To relate \eqref{52} to the standard change of measure technique,
consider the upper bound on $E^*(P)$ (which corresponds to a lower bound
on probabilities) in the case where the probabilities of the event of interest
are order 1 at the logarithmic scale, namely $E^*(Q)=0$. Then one can take
$\al\to1$. Since the divergence term converges (formally) to that
given in terms of the KL divergence, the standard change of measure method
recovers.

The bounds \eqref{52} are useful when for a given model of interest $P$,
one can find a reference model $Q$
for which the exponents are known or can be bounded, and at the same time,
one can efficiently estimate the divergence term. This is demonstrated in this
article several times. Whereas the case alluded to above,
in which both $P$ and $Q$ have i.i.d.\  structure, is most instructive,
we will apply the bounds \eqref{52} in scenarios that go far beyond that.
In fact, the bound we develop are more effective in situations where
the model of interest $P$ has long memory properties (such as, in the setting
of channel coding, models that have interference, fading or erasure with
long range correlations).

\subsection*{Second moment bounds}

A useful framework is when the true model consists of a small perturbation
of the reference model. Here we analyze a simple case where the alphabet is finite,
and obtain a bound involving the second moment of the perturbation size.
While the proof of the result is simple, it is an archetype of the argument
used several times in the sequel for more complicated models
in which the noise is dominant. These include the very noisy channel
(see p.\ 155, eq.\ (3.4.23) of \cite{VO})
$P(y|x)=Q(y)[1+\epsilon(x,y)]$ and, in the same spirit,
the weak interference channel
$P(y_t|x^{n-1},y^{t-1})=Q(y_t|x_t)[1+\epsilon(x^n,y^t)]$.

Let a vector $X^n$ take values in $\calY^n$ where $\calY$ is a finite set,
and assume that the vector is i.i.d.\  under both the measures
$P$ and $Q$. Denote by $P_n$ and $Q_n$ the respective probability laws of the vector.
Denoting $p=P_1$ and $q=Q_1$, assume that
$p(y)=q(y)[1+\epsilon(y)]$ for all $y\in\calY$,
where $\sum_y q(y)\epsilon(y)=0$. Assuming $q$ charges all of $\calY$,
so does $P$, provided that $\|\eps\|:=\max_y|\eps(y)|$ is small.
Let $\calA_n$ be any sequence of events of the form $\calA_n=\{X^n\in B_n\}$,
where $B_n$ is a Borel subset of $\R^n$ and use the notation \eqref{80}
for $E^*(P)$ and $E^*(Q)$.
\begin{proposition}\label{prop31}
Denote $\oo{\epsilon^2}=\sum_{y\in\calY}q(y)\eps^2(y)$. Then
\begin{equation}\label{83}
  E^*(P)\le
  \left(\sqrt{E^*(Q)}+\sqrt{\frac{\overline{\epsilon^2}}{2}}\right)^2+o(\|\eps\|^2).
\end{equation}
\end{proposition}

\proof
One has
\begin{eqnarray}
D_{\alpha}(q\|p)&=&\frac{1}{\al(\al-1)}\ln\left[\sum_{y}
q^{\alpha}(y)p^{1-\al}(y)\right]\\
&=&\frac{1}{\al(\al-1)}\ln\left[\sum_{y}
q(y)[1+\epsilon(y)]^{1-\al}\right]\\
&\le&\frac{1}{\al(\al-1)}\ln\left[\sum_{y}
q(y)[1+(1-\al)\epsilon(y)+\frac{1}{2}\al(\al-1)\epsilon^2(y)]+c(\al)\|\eps\|^3\right]\\
&=&\frac{1}{2}\sum_y q(y)\epsilon^2(y)+\tilde c(\al)\|\eps\|^3\\
&=&\frac{1}{2}\overline{\epsilon^2}+\tilde c(\al)\|\eps\|^3,
\end{eqnarray}
for suitable $c(\al)$ and $\tilde c(\al)$. Now, by the assumed i.i.d.\
structure, $D_\al(Q_n\|P_n)=nD_\al(q\|p)$. Thus by \eqref{52}, for every $\al>1$,
\begin{equation}
  E^*(P)\le\frac{\al}{\al-1}E^*(Q)+\al\Del^{Q,P}_\al
  \le\frac{\al}{\al-1}E^*(Q)+\al\Big[\frac{1}{2}\oo{\epsilon^2}+\tilde c(\al)\|\epsilon\|^3\Big].
\end{equation}
The function $\al\mapsto\al(\al-1)^{-1}u+\al v$, $\al\in(1,\iy)$, $u,v>0$
attains minimum at $\al^*=\sqrt{u/v}+1$ and the minimum is given by $(\sqrt u+\sqrt v)^2$.
Therefore
\begin{equation}
  E^*(P)\le\left(\sqrt{E^*(Q)}
  +\sqrt{\frac{\overline{\epsilon^2}}{2}}\right)^2+\al^*\tilde c(\al^*)\|\eps\|^3=
  \left(\sqrt{E^*(Q)}
  +\sqrt{\frac{\overline{\epsilon^2}}{2}}\right)^2+O(\|\eps\|^3).
\end{equation}
\qed

\begin{remark}
A more general setting is discussed in a recent paper \cite{ervhar2}
(specifically eq.\ (50) therein) where for a parametric family
$\{P_\theta,~\theta\in\Theta\}$, one has
\begin{equation}
\lim_{\theta'\to\theta}\frac{D_{\al}(P_\theta\|P_{\theta'})}{(\theta'-\theta)^2}=
\frac{J(\theta)}{2},
\end{equation}
where $J(\theta)$ is the Fisher information.
In this case, the bound \eqref{83} is valid with
$\sqrt{\frac{\overline{\epsilon^2}}{2}}$
replaced by $\sqrt{J(\theta)/2}\cdot|\theta'-\theta|$.
\end{remark}

\section{Applications to channel coding}\label{sec4}

This section addresses the use of the lower and upper bounds \eqref{52}
in the context of channel coding. We begin by considering, in Subsection
\ref{sec41}, a general framework
where we describe the relevance of the bounds in three contexts:
(1) Bounds on performance for a given channel in terms of a reference channel;
(2) Bounds for mismatched decoding;
(3) Robust bounds.
In Subsections \ref{sec42}--\ref{sec45},
we consider several specific channel models of interest,
where our methods yield new bounds.
These include interference with long range dependence, discrete and continuous
time Gaussian (and non-Gaussian)
channels with fading, and the binary channel with erasure.

\subsection{Generalities}
\label{sec41}

\subsubsection*{Setting and main estimates}

In channel coding, messages are encoded,
transmitted over a noisy channel and decoded. The precise setting
that we shall use is as follows.
A message $m$ from a set of $M=e^{nR}$ messages, $\calM=\{0,1,\ldots,M-1\}$,
is encoded into a codeword $\bx_m=(x_{m,1},\ldots,x_{m,n})$ of length $n$, whose
coordinates all take on values in a space $\calX$, that for the purposes
of this paper may be either finite or a Euclidean space
$\R^k$ (some $k\ge1$).
Here, $R> 0$ is the coding rate in nats per channel use.
We let $\calC_n=\{\bx_0,\bx_1,\ldots,\bx_{M-1}\}$ denote the codebook.
Our analysis allows for the codebook to be either deterministic or random,
settings which we refer to as deterministic and random coding, respectively.
When a codeword $\bx_m\in\calC_n$ is transmitted over a channel,
a channel output $\by=(y_1,\ldots,y_n)\in\calY^n$ is produced,
where again $\calY$ is either a given finite set or $\R^\ell$ (some $\ell\ge1$).
The decoder observes the vector $\by$ and produces an estimate
$\hat{m}\in\calM$ using a metric decoder, i.e.,
\begin{equation}
\hat{m}=\mbox{argmin}_{m\in\calM} d_n(\bx_m,\by)
\end{equation}
where ties are broken by an arbitrary deterministic rule,
and $d_n(\bx,\by)$ is an additive {\it decoding metric} function,
that is, it takes the form
\begin{equation}
d_n(\bx,\by)=\sum_{i=1}^n d(x_i,y_i),
\end{equation}
where $d:\calX\times\calY\to[0,\iy)$ is a given Borel measurable function.
To describe the model probabilistically, we consider now the
input and output of the channel as random variables, and write them
as $\bX$ and $\bY$. The message and estimated message
are also random now but still denoted $m$ and $\hat m$, respectively.
The collection of these random variables (for all values of $n$), as well
as the codebooks (in the case of random coding) are defined on a probability
space $(\Om,\calF,P)$.
The probabilistic elements and assumptions of the model are as follows:
\\ (i) $m$ is uniformly distributed over $\calM$. Consequently (assuming throughout
that all codewords are distinct),
\begin{equation}\label{40}
\Pi(\bx)\dfn P(\bX=\bx)=\left\{\begin{array}{ll}
\frac{1}{M} & \bx\in\calC_n\\
0 & \mbox{elsewhere,}\end{array}\right.
\end{equation}
for deterministic coding, and
\begin{equation}\label{41}
\Pi(\bx)\dfn P(\bX=\bx|\,\calC_n)=\left\{\begin{array}{ll}
\frac{1}{M} & \bx\in\calC_n\\
0 & \mbox{elsewhere,}\end{array}\right.
\end{equation}
in the case of random coding.
\\(ii) The model for the channel is described by the
conditional distribution of $\bY$ given $\bX$. This conditional distribution
is denoted by
\begin{equation}
P(\by|\bx)\dfn P\{\bY=\by|\bX=\bx\}.
\end{equation}
If we denote by $\calE_n=\{\hat m\ne m\}$ the error event then
the error probability is given by $P(\calE_n)$.
In the case of random coding, this can be written as
$P(\calE_n)=\bE_{P}[P(\calE_n|\calC_n)]$, which is interpreted as
the mean probability of error when averaged over codes.
The decoding metric $d_n$ is not assumed to be matched to the channel
(as is the case, for example, when $P(\by|\bx)$ takes the product form
$\prod_{i=1}^np(y_i|x_i)$ and $d(x,y)$ is proportional to $-\ln p(y|x)$).
We will sometimes assume (without essential loss of generality) that the given code
$\calC_n=\{\bx_0,\bx_1,\ldots,\bx_{M-1}\}$ is a constant composition code (CCC),
that is, all codewords have the same empirical distribution, which converges
to a given probability distribution
$\mu=\{\mu(x),~x\in\calX\}$ as $n\to\infty$.

A {\it reference channel} is another probability measure, $Q$, on $(\Om,\calF)$,
which models a (possibly) different channel.
In this work, we will always assume that, under $Q$, the distribution of the codes
(in the case of random coding) as well as the probability
of each codeword, is the same as under $P$; specifically, \eqref{40}
and \eqref{41} are valid with $P$ replaced by $Q$.

For deterministic coding, let $P_n$ and $Q_n$
denote the joint distribution of the two $n$-vectors
$(\bX,\bY)$ under $P$ and $Q$, respectively. It will be assumed that, given $n$,
the correspondence between $m$ and $\bX$ is one-to-one. As a result,
the error event is measurable with respect to the $\s$-field generated
by $(\bX,\bY)$. In the case of random coding, it is not natural to assume
that the correspondence alluded to above is always one-to-one. In this case
we use the same notation, $P_n$ and $Q_n$, to
denote the respective distributions of the quadruple
$(\calC_n,m,\bX,\bY)$. The error event is then measurable with respect
to the $\s$-field of this quadruple.
Thus by \eqref{30}, we have for every $n$ and every $\al>1$, the lower bound
\begin{equation}
\label{43}
\frac{1}{n}\ln P(\calE_n)\ge
\frac{\al}{n(\al-1)}\ln Q(\calE_n)
-\frac{\al}{n} D_\al(Q_n\|P_n),
\end{equation}
and the upper bound
\begin{equation}
\label{44}
\frac{1}{n}\ln P(\calE_n)\le
\frac{\al-1}{n\al}\ln Q(\calE_n)
+\frac{\al-1}{n} D_\al(P_n\|Q_n).
\end{equation}
Adapting the notation \eqref{80} to the present setting, we write
\begin{equation}
E_*(R,\hat P, d)=-\limsup_{n\to\iy}\frac{1}{n}\ln \hat P(\calE_n),
\qquad
E^*(R,\hat P, d)=-\liminf_{n\to\iy}\frac{1}{n}\ln \hat P(\calE_n),
\end{equation}
where $\hat P\in\calP$ is any channel model, and
we emphasize the dependence on the rate $R$ and on the metric $d$
(however, in the sequel, we sometimes suppress the dependence on $R$ and $d$ when
there is no room for confusion).
The notation $\Del_\al$ from \eqref{80+} will be used here with $\hat P_n$
and $\hat Q_n$ again denoting the respective joint distribution of the $n$-vectors
$(\bX,\bY)$. We thus obtain from \eqref{52}, for every $\al>1$, the bounds
\begin{align}\notag
  &\frac{\al-1}{\al}E_*(R,Q,d)-(\al-1)\Del^{P,Q}_\al\le
  E_*(R,P,d)\\ \label{45}
&\qquad \qquad \le E^*(R,P,d)\le\frac{\al}{\al-1}E^*(R,Q,d)+\al\Del^{Q,P}_\al.
\end{align}

\subsubsection*{Three interpretations of the bounds}

We identify three ways in which the above bounds can be used. In all cases, we
think of $P$ as the true channel model and $Q$ as a reference.

(i) {\it Bounds on performance of the true channel in terms of a reference channel.}

One can obtain lower [upper] bounds on error exponents for
true channel models by means of a lower [resp., upper] bound for a reference model.
Suppose that $d$ and a reference channel $Q$ are given, where $d$ is matched
to $Q$. More generally, suppose that a parametric family $\{Q_\theta\}$
is given such that a given, fixed metric $d$ is matched
to each member of the family. Assume further that one knows a lower bound,
$E_L(R,Q_\theta,d)$ on the error exponent $E(R,Q_\theta,d)$. Then for a metric
$d_P$ that is matched to $P$, we obtain
\begin{equation}\label{46}
  E_*(R,P,d_P)\ge E_*(R,P,d)\ge
  \sup_{\al>1}\sup_\theta\Big[
  \frac{\al-1}{\al}E_L(R,Q_\theta,d)-(\al-1)\Del^{P,Q_\theta}_\al\Big].
\end{equation}

Similarly, an upper bound is possible for given $P$ and $d$,
when for reference channels $Q$ one knows an upper bound $E_U(R,Q,d)$
on $E(R,Q,d)$ (when $d$ is not necessarily matched to $Q$) and then
\begin{equation}\label{47}
  E^*(R,P,d)\le\inf_{\al>1}\inf_\theta\Big[\frac{\al}{\al-1}E_U(R,Q,d)
  +\al\Del^{Q_\theta,P}_\al\Big].
\end{equation}

(ii) {\it Bounds on performance of mismatched decoding.}

When $d$ is matched to a reference channel $Q$, or a parametric family thereof,
the second inequality in \eqref{46} serves as an upper bound
on the mismatched error exponent (of using $d$ with the true channel $P$)
in terms of matched error exponent bounds (of using $d$ with the reference
channels $Q_\theta$ to which it is matched). A similar statement is
valid for the upper bound \eqref{47}.
To recapitulate, the above inequalities give bounds on the error exponents
under the true channel, operating with a decoder that is matched to
another channel in terms of error exponents of the latter.

(iii) {\it Robust bounds.}

Consider a family $F$ of true channels. As a performance criterion for
the decoder, it is of interest to study the minimum error exponent within
the family, namely
\begin{equation}
  \calE(R,F,d):=\inf_{P\in F}E(R,P,d).
\end{equation}
Optimizing over decoders gives
\begin{equation}\label{48}
  \calE(R,F):=\sup_{d}\calE(R,F,d).
\end{equation}
Thus $\calE(R,F)$ is the best possible guarantee on the performance
of all channels within the family when the communication system
operates with a single decoder $d$ (where `best' refers to the selection of $d$).
We can take advantage of the fact that the aforementioned bounds
for a fixed channel model, $P$, are independent of $P$ for $P\in F$,
in order to obtain information on $\calE(R,F)$.
To this end, fix a reference channel $Q$,
and assume that it is a member of the family $F$.
Then automatically, $\calE(R,F)\le E(R,Q,d_Q)$, where $d_Q$ is matched to $Q$.
As far as a lower bound is concerned, recall that for $P\in F$, and fixed $\al$,
\begin{equation}
E(R,P,d)\ge\frac{\al-1}{\al}E(R,Q,d)-(\al-1)\Del^{P,Q}_\al.
\end{equation}
Let $r(\al)=(\al-1)\sup_{P\in F}\Del^{P,Q}_\al$. Then, for $\al>1$,
\begin{equation}
\calE(R,F)\ge \frac{\al-1}{\al}\sup_d \calE(R,Q,d)-r(\al).
\end{equation}
Whereas the max-min problem posed by \eqref{48} is typically notoriously hard,
the optimization problem that now appears in the bound is easy to handle,
since the optimal decoder for $Q$ is the one matched to it.
Thus we have
\begin{equation}\label{59}
\sup_{\al>1}\Big[\frac{\al-1}{\al} E(R,Q,d_Q)-r(\al)\Big]\le \calE(R,F)\le E(R,Q,d_Q).
\end{equation}

\skp

The points of view (i)--(iii) presented above will be further explored and
demonstrated for the specific models to be considered.

\skp

\subsubsection*{Implications on general memoryless channels}

Here we consider the mismatched channel problem where both $P$ and $Q$
are memoryless. For simplicity, we assume that $\calX$ and $\calY$ are discrete.
Given a metric $d$, it is natural to consider as a
parametric family of reference channels $Q=Q_{\theta,\psi}$ given by
\begin{equation}
Q(\by|\bx)=\prod_{i=1}^n q(y_i|x_i),
\end{equation}
where
\begin{equation}
q(y|x)=q_{\theta,\psi}(y|x)=\frac{\psi(y)\cdot e^{-\theta
d(x,y)}}{\sum_{y'\in\calY}\psi(y')e^{-\theta d(x,y')}},
\qquad x\in\calX, y\in\calY,
\end{equation}
and $\theta \ge 0$ and $\psi(y)\ge 0$, $y\in\calY$, are the parameters of the channel.
Then the decoding metric $d$ is matched to each of these channels,
namely $d$ is the maximum likelihood (ML) decoding metric for $Q_{\theta,\psi}$
for each $\theta$ and $\psi$. It is instructive to note that,
as $\theta\to 0$, the channel
becomes ``noisier'', i.e., the output becomes proportional to $\psi(y)$, independently
of the input.
Assume a constant composition code.
Then, for $Q=Q_{\theta,\psi}$, we can calculate the divergence term as
\begin{align}
\frac{\al}{n} D_\al(Q_n\|P_n)
&=\frac{1}{n(\al-1)}
\ln\left(\sum_{\bx\in\calC_n}\sum_{\by\in\calY^n}[\Pi(\bx)Q(\by|\bx)]^\al
[\Pi(\bx)P(\by|\bx)]^{1-\al}\right)\\
&=\frac{1}{n(\al-1)}\ln\left(\sum_{\bx\in\calC_n}\Pi(\bx)\sum_{\by\in\calY^n}
\prod_{i=1}^n[q^\al(y_i|x_i)p^{1-\al}(y_i|x_i)]\right)\\
&=\frac{1}{n(\al-1)}\ln\left(\sum_{\bx\in\calC_n}\Pi(\bx)
\prod_{i=1}^n\left[\sum_{y\in\calY}q^\al(y|x_i)p^{1-\al}(y|x_i)\right]\right)\\
&=\frac{1}{n(\al-1)}\ln\left(\sum_{\bx\in\calC_n}\Pi(\bx)
\prod_{\bar x\in\calX}
\left[\sum_{y\in\calY}q^\al(y|\bar x)p^{1-\al}(y|\bar x)\right]
^{n\mu(\bar x)}\right)\\
&=\frac{1}{(\al-1)}\sum_{x\in\calX}\mu(x)\ln\left(
\sum_{y\in\calY}q^\al(y|x)p^{1-\al}(y|x)\right)\\
&=\al\sum_{x\in\calX}\mu(x)D_\al (q(\cdot|x)\|p(\cdot|x)).
\end{align}
Thus the term reduces to one that involves R\'enyi divergence at the single-letter
conditional marginals. Substituting in \eqref{45}, we obtain
\begin{equation}\label{02}
E(R,P,d)\le \frac{\al}{\al-1}E(R,Q,d)+\al\sum_{x\in\calX}
\mu(x)D_\al(q(\cdot|x)\|p(\cdot|x)).
\end{equation}
Now, $E(R,Q,d)$ is an error exponent for {\it matched decoding}
for the channel $Q$, and is therefore upper bounded by any upper bound
on the reliability function, such as the well-known straight--line bound
$E_{\mbox{\tiny sl}}(R,Q)$ (cf.\ Sections 3.6--3.8 of \cite{VO}).
Thus, we have
\begin{equation}
\label{bound}
E(R,P,d)\le \inf_{\psi,\theta}\inf_{\al> 1}\left[
\frac{\al}{\al-1} E_{\mbox{\tiny sl}}(R,Q_{\theta,\psi})+\al\sum_{x\in\calX}
\mu(x)D_\al(q_{\theta,\psi}(\cdot|x)\|p(\cdot|x))\right].
\end{equation}

\begin{remark}
To put \eqref{bound} in the context of known results,
let $I(\mu,Q)$ denote the single--letter mutual information between $X$ and $Y$,
induced by the joint distribution $\mu\times Q$, that is,
\begin{equation}
I(\mu,Q)=\sum_{x\in\calX}\mu(x)\sum_{y\in\calX}
q(y|x)\ln\left[\frac{q(y|x)}
{\sum_{x'\in\calX}\mu(x')q(y|x')}\right].
\end{equation}
In is known \cite{CK} that
\begin{equation}\label{85}
  E(R,P)\le\sup_{\mu}\inf_{Q:I(\mu,Q)\le R}D(Q\|P|\mu).
\end{equation}
Let us show that \eqref{bound} if fact reduces to \eqref{85}.
Given $R$, and a random coding distribution $\mu$, consider $Q$ for which $I(\mu,Q)\le R$.
Then $E_{\mbox{\rm\tiny sl}}(R,Q)=0$, and so
eq.\ (\ref{bound}) is further upper bounded by
\begin{equation}
\label{bound1}
E(R,P,d)\le \inf_{\al> 1}
\al\sum_{x\in\calX}
\mu(x)D_\al(q(\cdot|x)\|p(\cdot|x)).
\end{equation}
Now, $\alpha D_\al(q(\cdot|x)\|p(\cdot|x))$ is a
monotonically non--decreasing as a function of $\alpha$, and taking
the limit $\alpha\downarrow 1$ results in
$\sum_{x\in\calX}\mu(x)D(q(\cdot|x)\|p(\cdot|x))$, which recovers \eqref{85}
by minimizing over $Q$ and maximizing over $\mu$.
\end{remark}

\subsubsection*{Iterated use of the LPCB}
Recall the general upper bound \eqref{44}
\begin{equation}\label{60}
\frac{1}{n}\ln P(\calE_n)\le
\frac{\al-1}{n\al}\ln Q(\calE_n)
+\frac{\al-1}{n} D_\al(P_n\|Q_n),
\end{equation}
which holds for any pair of channel models $P$ and $Q$ and every $\al>1$.
We can iterate this estimate so as to compare another model, $\hat P$,
to $Q$ by relating it first to $P$ and $P$ to $Q$.
This may be useful in situations when estimating the divergence of
$P$ from $Q$ and that of $\hat P$ from $P$ is easier than estimating
the divergence of $\hat P$ from $Q$.
Indeed, expressing relation \eqref{60}
for $\hat Q$ and $P$ gives, for any $\beta>1$,
\begin{equation}\label{61}
\frac{1}{n}\ln\hat P(\calE_n)\le
\frac{\beta-1}{n\beta}\ln P(\calE_n)
+\frac{\beta-1}{n} D_\beta(\hat P_n\|P_n).
\end{equation}
Consequently, for any $\al>1$ and $\beta>1$,
\begin{equation}\label{62}
\frac{1}{n}\ln \hat P(\calE_n)\le
\frac{(\al-1)(\beta-1)}{n\al\beta}\ln Q(\calE_n)
+\frac{(\al-1)(\beta-1)}{n\beta} D_\alpha(P_n\|Q_n)
+\frac{\beta-1}{n} D_\beta(\hat P_n\|P_n).
\end{equation}
We use this approach in one of the results of Subsection \ref{sec42}.

\subsection{Interference with long range dependence}\label{sec42}

In this section we are interested in a channel of the form
\begin{equation}\label{63}
  Y_t=X_t+g_t(X^n,Y^{t-1})+W_t,
\end{equation}
for a generic sequence of functions $g_t:\R^n\times\R^{t-1}\to\R$.
Here $\{W_t\}$ is an i.i.d.\ $\calN(0,\sig^2)$ noise
(although we will also address a more general i.i.d.\ noise in the sequel).
The main assumption is that the interference functions $g_t$ are bounded.
However, no assumption is made about $g_t$ that limits the range of correlations
of the interference signal.
We let $P$ be a probability measure under which $\{W_t\}$ is as described above,
and $\{X_t\}$ and $\{W_t\}$ are mutually independent. This model will
be studied via the reference channel $Q$, under which
\begin{equation}\label{64}
  Y_t=X_t+\tilde W_t,
\end{equation}
where $\{\tilde W_t\}$ are i.i.d.\  $\calN(0,\sig^2/s)$ ($s>0$ being a parameter),
independent of $\{X_t\}$.
Thus under the true channel,
\begin{equation}\label{53}
P(\by|\bx)=(2\pi\sigma^2)^{-n/2}\exp\left\{-\frac{1}{2\sigma^2}
\sum_{t=1}^n[y_t-x_t-g_t(x^n,y^{t-1})]^2\right\},
\end{equation}
while under $Q=Q_s$,
\begin{equation}\label{54}
Q(\by|\bx)=\Big(\frac{2\pi \sigma^2}{s}\Big)^{-n/2}
\exp\left\{-\frac{s}{2\sigma^2}\sum_{t=1}^n(y_t-x_t)^2\right\}.
\end{equation}

\begin{theorem}\label{th31}
Denote by $E_{\mbox{\rm\tiny sl}}(R,Q_s)$ the straight--line (upper) bound
on $E(R,Q_s)$. Assume that for every $n$, and $t$,
\begin{equation}
\max_{x^n,y^{t-1}}|g_t(x^n,y^{t-1})|\le \Gam_t,
\end{equation}
and $\sum_{t=1}^n\Gam_t^2\le n\Gam^2$ for constants $\{\Gam_t\}$ and $\Gam$. Then,
for any sequence of codes and any decoder,
\begin{align}
E(R,P) \le \inf_{\al > 1}\inf_{s> 1-1/\al} & \Big\{\frac{\al
E_{\mbox{\tiny sl}}(R,Q_s)}{\al-1}+
\frac{\al\ln s}{2(\al-1)} \notag \\ \label{49}
&
-\frac{\ln[1+\al(s-1)]}{2(\al-1)}+\frac{\al s\Gam^2}{2\sigma^2[1+\al(s-1)]}\Big\}.
\end{align}
\end{theorem}

The proof of this result, that appears in the appendix,
uses the following identity in estimating the R\'enyi divergence.
For any real $u$, $v$, $a$ and $b$ such that $a+b>0$,
\begin{equation}
\label{identity}
\int_{-\infty}^{+\infty}\mbox{d}y\exp\{-a(y-u)^2-b(y-v)^2\}=\sqrt{\frac{\pi}{a+b}}
\cdot\exp\left\{-\frac{ab(u-v)^2}{a+b}\right\}.
\end{equation}
This identity is used, in addition, in several other proofs in the sequel.

We emphasize that for the model under consideration, the authors are not aware
of any other alternative bound on the error exponents.

\skp

Consider now the choice $s=1$ in \eqref{49}. In this case, the expression simplifies to
\begin{equation}
E(R,P)\le \inf_{\al > 1}\left\{\frac{\al E_{\mbox{\tiny sl}}(R,Q_1)}{\al-1}+
\frac{\al \Gam^2}{2\sigma^2}\right\}.
\end{equation}
The optimal $\alpha$ is easily found to be
\begin{equation}
\alpha^*=1+\frac{\sigma\sqrt{2E_{\mbox{\tiny sl}}(R,Q_1)}}{\Gam},
\end{equation}
which yields
\begin{equation}\label{500}
E(R,P)\le E_U(R)\dfn \left(\sqrt{E_{\mbox{\tiny
sl}}(R,Q_1)}+\frac{\Gam}{\sqrt{2}\sigma}\right)^2.
\end{equation}

The structure of the above bound is reminiscent of the bound
from Proposition \ref{prop31}. However, the bound above is
valid not only for weak interference. Results of similar structure appear
several times in the sequel.

Note that the above bound has a clear weakness of having a floor of $\Gam^2/2\sigma^2$
independent of the rate $R$.
This is an inherent limitation stemming from the way the we apply the bound.
However, one may apply additional considerations to address this difficulty.
Specifically, one can use the idea of the straight--line
bound (c.f.\ Theorem 3.8.1 in \cite{VO}\footnote{This theorem requires,
in principle,
the sphere--packing bound for list decoders, and for such a general channel,
we don't know the sphere--packing bound. Nonetheless, one can still use the theorem
when the higher rate is the capacity since the probability of list error is
bounded away from zero, for any codebook
of size
$M=e^{n(C+\lambda+\epsilon)}$ and list of size $e^{\lambda n}$, as can easily be shown by a simple
extension of Fano's inequality for list decoding. This is done by using the
fact that $H(\bX|\bY,~\mbox{no list error})\le n\lambda$ (unlike the case of
ordinary decoding where $H(\bX|\bY,~\mbox{no error})=0$).}),
to improve the bound using the smallest straight--line function that touches
the curve $E_U(R)$, passing through the point $(C,0)$, where $C$ is
the capacity of the true channel. The latter is upper bounded by $C\le
\frac{1}{2}\ln[1+(\sqrt{S}+\Gam)^2/\sigma^2]$, where $S$ is an upper bound on the
average power of $\bX$.
In what follows we will denote this improved bound by $E_1(R)$.

\subsubsection*{Very noisy channel}

We now focus on the case of a very noisy channel, where bounds can be computed explicitly
and insight can be obtained.
We thus study the implication of Theorem \ref{th31}, specifically of \eqref{500},
to the case where
$\sigma^2 \gg S+\Gam^2$, where $\{X_t\}$ satisfies
$\sum_{t=1}^nX_t^2\le nS$ a.s., for a given power limitation $S>0$. In this case,
the capacity of the reference channel (with $s=1$)
is about $C_Q=S/2\sigma^2$ and the capacity of the true channel is (upper bounded by)
$C=(\sqrt{S}+\Gam)^2/2\sigma^2$. The error exponent is given by
(see p.\ 157, eq.\ (3.4.33) of \cite{VO})
\begin{equation}
E(R,Q)=\left\{\begin{array}{ll}
C_Q/2 - R & R < C_Q/4\\
\left(\sqrt{C_Q}-\sqrt{R}\right)^2 & C_Q/4 \le R < C_Q\\
0 & R > C_Q.\end{array}\right.
\end{equation}
Now, accordingly,
\begin{eqnarray}
E_U(R)&=&\left\{\begin{array}{ll}
\left[\sqrt{C_Q/2 - R}+\Gam/(\sqrt{2}\sigma)\right]^2 & R < C_Q/4\\
\left[\sqrt{C_Q}+\Gam/(\sqrt{2}\sigma)-\sqrt{R}\right]^2 & C_Q/4 \le R < C_Q\\
\Gam^2/(2\sigma^2) & R > C_Q\end{array}\right.\nonumber\\
&=&\left\{\begin{array}{ll}
\left[\sqrt{C_Q/2 - R}+\Gam/(\sqrt{2}\sigma)\right]^2 & R < C_Q/4\\
\left(\sqrt{C}-\sqrt{R}\right)^2 & C_Q/4 \le R < C_Q\\
\Gam^2/(2\sigma^2) & R > C_Q.\end{array}\right.
\end{eqnarray}

Note that at least in the intermediate range, between $C_Q/4$ and $C_Q$, the
bound is tight in the sense that there exists an interference signal that
achieves it. It corresponds to the coherent sum of the desired signal and the
interference, which is the case when $g_t(x^n,y^{t-1})$ is proportional to $x_t$.

The improvement at high rates is provided by the straight--line that passes through
the points $(C_Q,\Gam^2/2\sigma^2)$ and $(C,0)$. The result we thus obtain
for the very noisy channel is
\begin{equation}
E(R,P)\le E_1(R)\dfn\left\{\begin{array}{ll}
\left[\sqrt{C_Q/2 - R}+\Gam/(\sqrt{2}\sigma)\right]^2 & R < C_Q/4\\
\left(\sqrt{C}-\sqrt{R}\right)^2 & C_Q/4 \le R < C_Q\\
\frac{\Gam^2(C-R)}{2\sigma^2(C-C_Q)} & C_Q\le R < C.\end{array}\right.
\end{equation}

\begin{remark}
At rate zero (and general SNR),
the bound one obtains from the discussion above, by selecting $s=1$, is
\begin{equation}\label{56}
E_1(0)=\Big(\sqrt{E_{\mbox{\tiny ex}}(0,Q)}+\frac{\Gam}{\sqrt{2}\sigma}\Big)^2
=\Big(\sqrt{\frac{C_Q}{2}}+\frac{\Gam}{\sqrt 2 \sigma}\Big)^2
=\frac{(\sqrt{S}+\Gam\sqrt{2})^2}{4\sigma^2}.
\end{equation}
It turns out that for $R=0$ one can solve the full optimization problem \eqref{49},
including the minimization over the parameter $s$.
In fact, one can even solve an extended problem,
in which the reference model has one additional
free parameter, namely a gain factor $\phi$: Instead of \eqref{54}, one considers
$Q=Q_{\theta,\phi}$ of the form $\prod_{t=1}^n Q(y_t|x_t)$, where
\begin{equation}\label{71}
Q(y|x)=\Big(\frac{s}{2\pi\sigma^2}\Big)^{1/2}\exp\Big\{-\frac{s}{2\sigma^2}(y-\phi x)^2\Big\},~~~
s> 0,~\phi> 0.
\end{equation}
However, the bound one obtains is exactly \eqref{56}.
\end{remark}

\subsubsection*{Lower bound on the exponent}

We can also derive a lower bound by appealing to \eqref{47}.
In this context, it is more natural to consider the setting of random coding
because existing bounds for reference models are of this type.
Denoting the sequence of random codes by $\{\calC_n\}$, the relevant divergence term
for using \eqref{47} is
\begin{equation}
  D_\al(P_n\|Q_n)=\frac{1}{\al(\al-1)}\ln
  \bE_Q\Big[\Big(\frac{P(\calC_n,m,\bX,\bY)}{Q(\calC_n,m,\bX,\bY)}\Big)^\al\Big].
\end{equation}
Recalling our assumption that
under the true model $P$ and under the reference model $Q$ the distribution
of the codes is equal, we have
\begin{equation}
  D_\al(P_n\|Q_n)=\frac{1}{\al(\al-1)}\ln
  \bE_Q\Big[\Big(\frac{P(\bX,\bY|\calC_n,m)}{Q(\bX,\bY|\calC_n,m)}\Big)^\al\Big].
\end{equation}
Now, the estimate on the divergence term appearing in the proof of
Theorem \ref{th31} can be carried out for the above in a similar manner,
and one obtains the same bound \eqref{57} regardless of the code $\calC_n$.
For simplicity, we can specialize to $s=1$, which gives
\begin{eqnarray}
E(R,P)&\ge& E_L(R,P)\dfn
\sup_{\al> 1}\left[\frac{\al-1}{\al}E(R,Q_1)-\frac{(\al-1)\Gam^2}{2\sigma^2}
\right],
\end{eqnarray}
the solution of which is
\begin{equation}\label{58}
E_L(R,P)=\left\{\begin{array}{ll}
\left(\sqrt{E(R,Q_1)}-\frac{\Gam}{\sqrt{2}\sigma}\right)^2 & E(R,Q_1)\ge
\Gam^2/2\sigma^2\\
0 & \mbox{elsewhere.}\end{array}\right.
\end{equation}

One can use the above bound to estimate the capacity of the the channel $P$.
It is bounded below by the
rate $R$ at which $E(R,Q_1)=\frac{\Gam^2}{2\sigma^2}$.
For the example of the very noisy channel,
this gives $C_P\ge(\sqrt{S}-\Gam)^2/2\sigma^2$.
This bound is attained by the interference signal that is anti--coherent with the desired
signal, i.e., $g_t(x^t,y^{t-1})=-\Gam x_t/\sqrt{S}$.

\subsubsection*{Robust bound interpretation}

All three interpretations mentioned in Subsection \ref{sec41} are relevant
for the results of this section. Specifically, the bounds
of Theorem \ref{th31} and \eqref{58}
are valid whether $d$ is matched to $P$ or not.
Next, to demonstrate the
robust bounds interpretation in the context of these results,
let $Q=Q_1$ denote the reference channel (with $s=1$) and for a fixed $\Gam$,
denote by $F$ the family of true channels $P$ for which $g_t$ are all bounded by $\Gam$.
Then by \eqref{59} and the bound
$r(\al)=(\al-1)\sup_{P\in F}\Del^{P,Q}_\al\le(\al-1)\Gam^2/(2\sigma^2)$
that follows from the previous paragraph, we have
\begin{equation}
\left(\sqrt{E(R,Q_1)}-\frac{\Gam}{\sqrt{2}\sigma}\right)^2
\le \sup_d \inf_{P\in F}E(R,P,d)\le E(R,Q_1).
\end{equation}
Specifically, the performance of a single decoder, namely
the one matched to $Q_1$, is bounded by the above two bounds whenever
the interference signal is bounded by the constant $\Gam$.

\subsubsection*{Non Gaussian noise}

Here we use the idea of iterating the bound, as presented
in Subsection \ref{sec41}, in order address non-Gaussian i.i.d.\  noise.
Going back to the general setting of Theorem \ref{th31},
recall from \eqref{63} and \eqref{64} that under $P$ and $Q$, respectively, we
have the models
\begin{equation}\label{78}
  Y_t=X_t+g_t(X^n,Y^{t-1})+W_t,
\end{equation}
\begin{equation}
  Y_t=X_t+\tilde W_t,
\end{equation}
where $\{W_t\}$ and $\{\tilde W_t\}$ are i.i.d.\  $\calN(0,\sigma^2)$, independent of
$\{X_t\}$. Consider now an additional model $\hat P$ described by
\begin{equation}\label{79}
  Y_t=X_t+g_t(X^n,Y^{t-1})+\hat W_t,
\end{equation}
where $\{\hat W_t\}$ are i.i.d.\  but need not be Gaussian.
The main point is that estimating the divergence of $\hat P$ from $P$
is simple, whereas the estimates on the divergence
of $P$ from $Q$ have already been established, thus by appealing to
\eqref{62}, one can relate
$\hat P$ to $Q$ by combining the two estimates.

Denote by $\del(\beta)=D_\beta(\calL_{\hat P}(\hat W_1)\|\calL_P(W_1))$
the single--letter R\'enyi divergence, where for a measure $\mu$ and r.v.\ $U$,
$\calL_\mu(U)$ denotes the probability law of $U$ under $\mu$.
Since both $\{\hat W_t\}$ and $\{W_t\}$ are i.i.d., we can make use of the simple
fact that
\begin{equation}
  D_\beta(\calL_{\hat P}(\hat W^n)\|\calL_P(W^n))
  =n D_\beta(\calL_{\hat P}(\hat W_1)\|\calL_P(W_1))=n\del(\beta).
\end{equation}
Moreover,
since under both $\hat P$ and $P$, the noise sequence is independent of the signal
$\{X_t\}$ and the latter has the same law, it follows that
\begin{equation}
  D_\beta(\calL_{\hat P}(X^n,\hat W^n)\|\calL_P(X^n, W^n))=n\del(\beta).
\end{equation}
Now, denote by $\hat P_n$ and $P_n$ the respective laws of $(X^n,Y^n)$ under
$\hat P$ and $P$.
Note by \eqref{78} and \eqref{79} that
$(X^n,Y^n)=F_n(X^n,W^n)$ and $(X^n,Y^n)=G_n(X^n,\hat W^n)$
for suitable deterministic functions $F_n$ and $G_n$. As a result,
the data processing inequality (see Theorem 9 of \cite{ervhar})
gives $D_\beta(\hat P_n\|P_n)\le
D_\beta(\calL_{\hat P}(X^n,\hat W^n)\|\calL_P(X^n, W^n))$.
Hence
\begin{equation}\label{65}
  D_\beta(\hat P_n\|P_n)\le n\delta(\beta).
\end{equation}
Using \eqref{65} in \eqref{62} gives
\begin{equation}
  E(R,\hat P)\ge \frac{(\al-1)(\beta-1)}{\al\beta}E(R,Q)
  -\frac{(\al-1)(\beta-1)}{\beta}\Del_\al^{P,Q}-(\beta-1)\del(\beta).
\end{equation}
We use our previous results that estimate $\Del^{P,Q}_\al$ and optimize over
$\al$. With $E_L$ given by \eqref{58}, we have
\begin{equation}\label{66}
  E(R,\hat P)\ge \frac{(\beta-1)}{\beta}E_L(R,P) -(\beta-1)\del(\beta).
\end{equation}
An analogous estimate can be established for an upper bound on the exponent,
as well as for all other channel models that we treat in the sequel.

\begin{example}
Consider truncated Gaussian noise distribution for $\hat W_1$, namely, for a given constant
$u$, assume $f_{\hat W_1}(w)=z^{-1}f(w)1_{[-u,u]}(w)$, where $f(w)=(2\pi)^{-1/2}e^{-w^2/2}$
is the standard normal density, and $z=\int_{-u}^uf(w)\dd w$.
Assume $W_1$ is standard normal. It is easy to see that
\begin{equation}
\del(\beta)=D_\beta(\hat W_1\|W_1)=\frac{\ln(1/z)}{\beta}.
\end{equation}
Thus using \eqref{66} and taking the limit $\beta\to\iy$,
\begin{equation}
  E(R,\hat P)\ge E_L(R,P)+\ln z.
\end{equation}
\end{example}

\subsubsection*{Robust bounds for the ISI channel}

We next study the Gaussian intersymbol interference (ISI) channel model, denoted
by $P$, given by
\begin{equation}\label{75}
  Y_t=X_t+\sum_{i=1}^kh_ix_{t-i}+W_t,
\end{equation}
where $\{W_t\}$ is i.i.d.\  $\calN(0,\sigma^2)$, independent of $\{X_t\}$,
and $\bh=(h_1,\ldots,h_k)^T$ is given.
While the proposed method yields new results for interference with
unlimited correlation length (Theorem \ref{th31}),
an analogous treatment of the model \eqref{75} turns out not to be useful,
as it leads to bounds
that are inferior to existing bounds, for both matched and mismatched decoding.
However, as we now demonstrate, the robust bound interpretation discussed
in Subsection \ref{sec41} gives rise to new results for this model.

Note that the model is a special case of the main model studied in this section.
Because of the special structure of the interference \eqref{75} and some further
assumptions we make regarding the correlation
structure, the bounds that we are able to provide are much more explicit than
those given by Theorem \ref{th31}.

The following will be assumed.
The rate is $R=0$, and the channel is very noisy, that is, $\sigma\gg\sqrt{S}$.
The decoder uses the mismatched decoding metric $d(x,y)=(x-y)^2$.
All codewords have energy
\begin{equation}\label{73}
\sum_{t=1}^n x_t^2=nS
\end{equation}
and a fixed empirical autocorrelation structure
\begin{equation}\label{74}
\sum_{t=i+1}^n x_tx_{t-i}=nc_iS.
\quad i=1,2,\ldots,k.
\end{equation}
Denote $\bc=(c_1,\ldots,c_k)^T$ and $\bC=[c_{|i-j|}]_{i,j=1}^k$ and let
$r_1=\bh^T\bc$ and $r_2=\bh^T\bC\bh$.
Then $r_1$ and $r_2$ are related to the empirical
correlation between signal and interference
$g_t:=\sum_{i=1}^kh_ix_{t-i}$ and interference power, respectively. Specifically,
\begin{equation}
\sum_{t=1}^nx_tg_t=nSr_1,\qquad \sum_{t=1}^ng_t^2=nSr_2.
\end{equation}
Note that always $r_2\ge r_1^2$.

\begin{theorem}\label{th45}
Consider a sequence of codes satisfying \eqref{73} and \eqref{74}
for a specific vector $\bc$. Denote by $F$ the family of
true models $P$ of the form \eqref{75}, where $h$ varies over all vectors having fixed
$r_1$ and $r_2$. Denote $a=r_2-r_1^2$ and $b=(1+r_1)^2$. If $a<b$ then
\begin{equation}\label{81}
  \frac{S}{4\sigma^2}(\sqrt b-\sqrt a)^2
  \le
  \sup_d\inf_{P\in F}E(0,P,d)
  \le\frac{S}{4\sigma^2}(\sqrt b+\sqrt a)^2.
\end{equation}
\end{theorem}
The proof appears in the appendix.

\subsection{Discrete time Gaussian channel with fading}

We consider the channel
\begin{equation}
Y_t=(1+\theta_t)X_t+ W_t,
\end{equation}
where $\{W_t\}$ is an additive noise process and $\{\theta_t\}$ is a fading process.
We let $P$ be a probability measure under which the processes
$\{\theta_t\}$, $\{W_t\}$ and $\{X_t\}$ are mutually independent, and
$\{W_t\}$ is i.i.d.\ $\calN(0,\sigma^2)$. Also, $\{X_t\}$
is assumed to satisfy the constraint $|X_t|\le \aaa$ for all $t$.
As a reference, consider a channel with no fading.
That is, consider a probability measure $Q$ under which
\begin{equation}
Y_t=X_t + \tilde W_t,
\end{equation}
where the law of
triplet $(X, \theta, \tilde W)$ under $Q$ is the same as that
of $(X, \theta, W)$ under $P$. In particular, under $Q$,
$\{\tilde W_t\}$ are i.i.d.\  $\calN(0,\sigma^2)$, and the three processes
$\{X_t\}$, $\{\theta_t\}$ and $\{\tilde W_t\}$ are mutually independent.

We assume that $\{\theta_t\}$ is a stationary, zero-mean Gaussian process and
that $r_k=E[\theta_0\theta_k]$ are absolutely summable.
Let $\Sig_\theta$ denote the spectral density of $\theta$, namely
\begin{equation}
  \Sig_\theta(\om)=\sum_{k=-\iy}^\iy r_ke^{-ik\om}.
\end{equation}

\begin{theorem}
  \label{th42}
  Let $P$ and $Q$ stand for the discrete-time Gaussian noise channel with
  and, respectively, without fading, described above.
  Denote $c=c(\al)=\al(\al-1)\aaa^2/(2\sigma^2)$. Then for any $\al>1$ such that
  $2c\sup_\om \Sig_\theta(\om)<1$,
\begin{align}
E(P)\le\frac{\al}{\al-1}E(Q)
-\frac{1}{4\pi(\al-1)}\int_0^{2\pi}\ln[1-2c\Sig_\theta(\om)]\dd\om,
\end{align}
\begin{equation}
E(P)\ge\frac{\al-1}{\al}E(Q)
+\frac{1}{4\pi\al}\int_0^{2\pi}\ln[1-2c\Sig_\theta(\om)]\dd\om.
\end{equation}
\end{theorem}
See the appendix for a proof.

Note that for fixed $\al$, the gap between the upper and lower bound increases with
$\Sig_\theta$. This occurs due to the fact that the distance between the model
$P$ and the reference model $Q$, as measured in terms of the divergence,
increases by strengthening the fading. When $\Sig_\theta\equiv0$, the models
$P$ and $Q$ agree, and then so do the upper and lower bounds (upon optimizing
over $\al$).

While it is difficult to optimize over $\al$ in general, in the next paragraph
we consider special cases where the results are more explicit.

\subsubsection*{AR fading model}
Consider the case of $\{\theta_t\}$ given by the autoregressive (AR) model
\begin{equation}
\theta_t=a\theta_{t-1}+b\hat W_t,
\end{equation}
where $\{\hat W_t\}$ are i.i.d.\ $\calN(0,1)$, $|a|<1$ and $\{\theta_t\}$ is stationary.
We have $r_k=r_0a^{|k|}$, $r_0=b^2/(1-a^2)$, and
\begin{equation}
\Sig_\theta(\om)=\frac{b^2}{1-2a\cos(\om)+a^2}.
\end{equation}
Thus $f(\om) \dfn
 1-2c(\al)\Sig_\theta(\om)$ gives
\begin{equation}
f(\om)=\frac{1-2a\cos(\om)+a^2-2cb^2}{1-2a\cos(\om)+a^2},
\end{equation}
and so, whenever $f$ is bounded away from zero, which holds iff
\begin{equation}\label{21}
(1-|a|)^2>2c(\al)b^2,
\end{equation}
one has
\begin{equation}\label{20}
E(P)\le\frac{\al}{\al-1}E(Q)-\frac{1}{4\pi(\al-1)}\int_0^{2\pi}
\ln f(\om)\dd\om.
\end{equation}

We next further develop \eqref{20} based on the residue theorem,
by which one has $\int_0^{2\pi}\ln(1+re^{i\om})\dd\om=0$
whenever $|r|<1$ for the complex logarithmic function $\ln(\cdot)$.
Specifically, if we express $f$ as
\begin{equation}
f(\om)=k\frac{(1-re^{-i\om})(1-re^{i\om})}{(1-ae^{-i\om})(1-ae^{i\om})},
\end{equation}
for a real $r$ with $|r|<1$, and $k>0$, then $\int_0^{2\pi}\ln f(\om)\dd\om=2\pi\ln k$.
To calculate $r$ and $k$, write for $z\in\C$,
\begin{equation}
  (1-az)(1-az^{-1})-2cb^2=k(1-rz)(1-rz^{-1}).
\end{equation}
The solution to this is $k=a/r$,
\begin{equation}
  r_{1,2}=\frac{\xi(\al)\pm\sqrt{\xi(\al)^2-4a^2}}{2a}.
\end{equation}
where
\begin{equation}
  \xi(\al)=1-2c(\al)b^2+a^2.
\end{equation}
Under \eqref{21}, the discriminant is positive, and therefore
$r_{1,2}$ and $k$ are real numbers. Also, one checks that under
\eqref{21}, $r_1>1$ hence not to be considered.
As for $r_2$, we have $|r_2|<1$ under that condition. Thus $k=a/r_2$, and we have
\begin{align}
  E(P)&\le\frac{\al}{\al-1}E(Q)-\frac{2\pi\ln k}{4\pi(\al-1)}\\
%  &=\frac{\al}{\al-1}E(Q)+\frac{1}{2(\al-1)}
%  \ln\frac{1-2cb^2+a^2-\sqrt{(1-2cb^2+a^2)^2-4a^2}}{2a^2}\\
  \label{22}
  &=\frac{\al}{\al-1}E(Q)+\frac{1}{2(\al-1)}
  \ln\frac{\xi(\al)-\sqrt{\xi^2(\al)-4a^2}}{2a^2}.
\end{align}
%where the last line uses the notation
%$c=\al(\al-1)p$, $p=\aaa^2/(2\sigma^2)$.

The limit case $b\to0$: This is when the fading amplitude goes to zero,
we have the bound converging to $\al/(\al-1)E(Q)$, and optimizing over $\al$
gives $E(Q)$, that is the best possible bound under the circumstances.

Using the lower bound gives the following
bound, complementing \eqref{20}, namely
\begin{align}
  E(P)&\ge \frac{\al-1}{\al}E(Q)+\frac{1}{4\pi\al}
  \int_0^{2\pi}\ln[1-2c\Sig_\theta(\om)]\dd\om\\
  &=\frac{\al-1}{\al}E(Q)-\frac{1}{2\al}
  \ln\frac{\xi(\al)-\sqrt{\xi^2(\al)-4a^2}}{2a^2},
  \label{25}
\end{align}
for all $\al>1$ satisfying \eqref{21}.

Figure 1 depicts the above bound as a function of $\alpha$ for various
values of $E(Q)$. Note that the range of the parameter $\alpha$
is of the form $(1,\alpha^*)$, where $\al^*$ is the smallest $\al$ which violates
condition \eqref{21}. The right end of the graphs
in Figures 1(a) and 1(b) correspond to $\al^*$.

\begin{figure}
\begin{center}
\includegraphics[width=35em,height=15em]{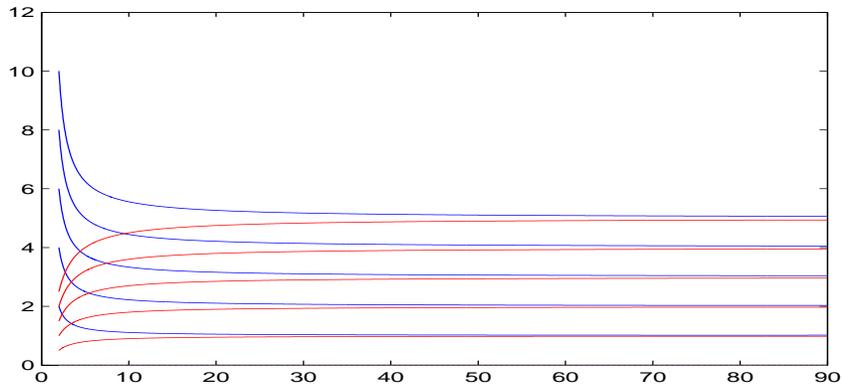}
\\
(a)
\\
\includegraphics[width=35em,height=15em]{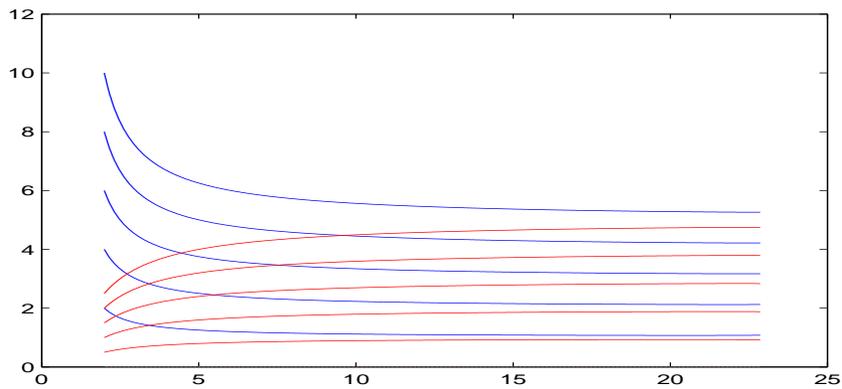}
\\
(b)
\caption{\sl\small
Plots of the upper bound \eqref{22} (blue) and lower bound \eqref{25} (red)
on $E(P)$ as function of $\alpha$.
The five graphs correspond to $E(Q)=1,2,3,4$ and $5$, with
$a=0.2$ and $\aaa^2/(2\sig^2)=0.1$,
where in plot (a), $b=0.02$, and in plot (b), $b=0.08$.
}
\end{center}
\end{figure}

We comment that the bounds are tight in the small fading limit. Namely,
as the amplitude of the fading perturbation goes to zero,
the optimal bounds (obtained by choosing $\al$ suitably) converge to $E(Q)$.
Indeed, as $b\to0$, the argument of the logarithmic function converges to $1$,
by which that follows.

Note that one can treat the small fading limit in greater generality (beyond the AR process).
Denote $\sigmax=\sup_\om \Sig_\theta(\om)$. Fix $0<\del<1$ and assume $2c\sigmax\le\del$.
Denote $\kappa=\frac{1}{2(1-\del)^2}$.
Using the bound $\ln(1+x)\ge x-\kappa x^2$ for all $x$ s.t.\ $|x|<\del$ in Theorem
\ref{th42} gives, for every fixed $\al>1$,
\begin{align}
E(P)
&\le\frac{\al}{\al-1}E(Q)+\frac{1}{4\pi(\al-1)}
\int_0^{2\pi}[2c\Sig_\theta(\om)+\kappa\del^2]\dd\om
\\
&=\frac{\al}{\al-1}E(Q)+\frac{\al \aaa^2 r_0}{2\sig^2} + \frac{\kappa\del^2}{2(\al-1)}.
\label{84}
\end{align}
Optimizing over the parameter $\al$ in the range
$\{\al>1\}\cap\{2c(\al) \sigmax \le\del\}$ can now be carried out easily
(in a manner similar to that in Proposition \ref{prop31}).

\subsection{Continuous--time white noise with fading}

A standard model for a white Gaussian channel in continuous time is given by
\begin{equation}\label{37}
  Y_t=\int_0^tX_s\dd s+\sigma W_t,
\end{equation}
where $\{W_t\}$ is a Brownian motion. Let $Q$ be a probability measure
under which $\{W_t\}$ is a standard Brownian motion, and let
$\{X_t\}$ and $\{\theta_t\}$ be real-valued processes
such that the three processes $\{W_t\}$, $\{X_t\}$ and $\{\theta_t\}$ are
mutually independent.
Assume that $\{X_t\}$ satisfies the amplitude constraint
$|X_t|\le\aaa$ for all $t$, $Q$-a.s., where $\aaa$ is a constant.
One can obtain from $Q$ a model for a channel with fading, in which $\theta$
is the fading process, by means of a change of measure.
To this end, consider the filtration
\begin{equation}
  \calF_t=\sigma\{X_s,\theta_s,W_s:s\in[0,t]\},
\end{equation}
and let
\begin{equation}
  Z_t=\exp\Big[\frac{1}{\sigma}\int_0^t\theta_sX_s\dd W_s-\frac{1}{2\sigma^2}
  \int_0^t(\theta_sX_s)^2\dd s\Big], \qquad t\ge0.
\end{equation}
It is assumed throughout that, for every $T>0$,
\begin{equation}\label{28}
\bE_{Q}\exp\Big\{\frac{\aaa^2}{2\sigma^2}\int_0^T\theta_s^2\dd s\Big\}<\iy.
\end{equation}
We later provide a sufficient condition for this to hold.
Note that, as a result, one has
$\bE_{Q}\exp\Big\{\frac{1}{2\sigma^2}\int_0^T(X_s\theta_s)^2\dd s\Big\}<\iy$, and so
Novikov's condition for $\{Z_t\}$ to be an $\{\calF_t\}$-martingale under
$Q$ is satisfied (see Corollary 3.5.13 of \cite{kar-shr}).
For $T>0$, let $Q_T$ and $P_T$ be probability measures on $\calF_T$,
defined by
\begin{equation}
  Q_T(\calA)=Q(\calA),\qquad P_T(\calA)=\bE_Q[1_\calA Z_T],\qquad \calA\in\calF_T,
\end{equation}
where $1_\calA$ denotes the indicator function of $\calA$.
Then $\frac{\dd P_T}{\dd Q_T}=Z_T$, and
by Girsanov's theorem (Theorem 3.5.1 of \cite{kar-shr}) one has
\begin{equation}\label{38}
  Y_t=\int_0^t(1+\theta_s)X_s\dd s+\sigma\tilde W_t,
\end{equation}
where, under $P_T$, the triplet $(\theta_t,X_t,\tilde W_t,t\in[0,T])$
has the same law as that of
$(\theta_t,X_t,W_t, t\in[0,T])$ under $Q_T$ (thus under $Q$).
In particular, under the measure $P_T$,
$\{\tilde W_t\}$ is a standard Brownian motion, and the three processes
$\{W_t\}$, $\{X_t\}$ and $\{\theta_t\}$ are mutually independent. As a result,
$P_T$ is a model for
an additive white Gaussian noise channel with a fading process $\{\theta_t\}$.

It is assumed that $\{\theta_t\}$ is a separable,
zero-mean stationary Gaussian process (under $Q_T$; equivalently under $P_T$).
The spectral density of $\{\theta_t\}$, that is, the function $\Sig_\theta$ for which
$\bE[\theta_0\theta_t]=\int_{-\iy}^\iy e^{it\om}\Sig_\theta(\om)\dd\om$,
is assumed to satisfy $\sigmax:=\esssup \Sig_\theta<\iy$.

The following, that can be seen as a continuous-time analogue of Theorem \ref{th42},
is the main result of this subsection.
\begin{theorem}\label{th43}
Let $P$ and $Q$ stand for the continuous time white noise channel
models with and without fading, described above.
Assume $p:=\aaa^2/(2\sigma^2)<1/(4\pi\sigmax)$.
Then \eqref{28} holds. Moreover,
with $c(\al)=\al(\al-1)p$,
for any $\al>1$ such that $c(\al)<1/(4\pi\sigmax)$,
\begin{equation}\label{86}
  E(P)\le\frac{\al}{\al-1}E(Q)-\frac{1}{4\pi(\al-1)}\int_{-\iy}^\iy
  \ln[1-4\pi c(\al)\Sig_\theta(\om)]\dd\om
\end{equation}
\begin{equation}\label{87}
  E(P)\ge\frac{\al-1}{\al}E(Q)+\frac{1}{4\pi\al}\int_{-\iy}^\iy
  \ln[1-4\pi c(\al)\Sig_\theta(\om)]\dd\om.
\end{equation}
\end{theorem}
See the appendix for a proof.

\skp

For an encoder/decoder optimized for $Q$, an expression for $E(R,Q)$
is well known (see Section 8.2 of \cite{gal}), namely, with $C=\aaa^2/(2\sigma^2)$,
\begin{equation}
  E(R,Q)=\begin{cases}
    C/2-R & R<C/4\\
    (\sqrt C-\sqrt R)^2 & C/4\le R<C\\
    0 & R\ge C.
  \end{cases}
\end{equation}
As a result, \eqref{86} and \eqref{87} give bounds on the
mismatched error exponents for the model with fading, when the encoder
and decoder are matched to $Q$. The lower bound \eqref{87} appears to be
new even for the matched channel exponent, that is, when the
right-hand side of \eqref{87}
serves as a lower bound on the error exponent for an encoder/decoder
that are matched to $P$.

\subsubsection*{Low frequency fading}
The expression in \eqref{29}
is simple when $\Sig_\theta$ is constant on its support. Specifically,
consider the case $\Sig_\theta(\om)=\Sig_0$ on the interval $[-B,B]$. Then
\begin{equation}
  \frac{\al-1}{\al}E(Q)-\frac{1}{\al}r(\al)\le
  E(P)\le\frac{\al}{\al-1}E(Q)+\frac{1}{\al-1}r(\al),
\end{equation}
where
\begin{equation}
  r(\al)=-\frac{2B}{4\pi}\ln[1-4\pi \al(\al-1)p\Sig_0],
\end{equation}
provided $\max\{\al(\al-1),1\}<1/(4\pi p\Sig_0)$.

\subsubsection*{Ornstein-Uhlenbeck fading}
Next consider a model where the fading process
takes the form of a stationary Ornstein-Uhlenbeck process, namely
\begin{equation}
  \dd\theta_t=-a\theta_t\dd t+b\,\dd\hat W_t,
\end{equation}
where $\hat W$ is a standard Brownian motion and $a>0$ and $b>0$ are constants.
Then the spectral density is given by $\Sig_\theta(\om)=(1/\pi)b^2/(a^2+\om^2)$,
and by a calculation from p.\ 130 of \cite{bry-dem}, one has
\begin{equation}
  -\frac{1}{4\pi}\int_{-\iy}^\iy\ln[1-4\pi c\Sig_\theta(\om)]\dd\om
  =\frac{1}{2}a-\frac{1}{2}\sqrt{a^2-4b^2c}
\end{equation}
provided $c<a^2/(4b^2)$.
Thus
\begin{equation}\label{35}
\frac{\al-1}{\al}E(Q)-\frac{1}{\al}r(\al)\le
  E(P)\le\frac{\al}{\al-1}E(Q)+\frac{1}{\al-1}r(\al)
\end{equation}
where
\begin{equation}
  r(\al)=\frac{1}{2}a-\frac{1}{2}\sqrt{a^2-4b^2p\al(\al-1)},
\end{equation}
provided $c(\al)=p\al(\al-1)<a^2/(4b^2)$ and $p<a^2/(4b^2)$.

While it is hard to optimize over $\al$, it is possible to do so
if we bound $r$ from above by
\begin{equation}
  \bar r(\al)=\frac12 a-\frac12\sqrt{a^2-4b^2p\al^2}
\end{equation}
and assume $p\al^2\le a^2/(4b^2)$. That is, $\al\in(1,a/(2b\sqrt{p}))$.
In particular, we must assume $a>2b\sqrt p$.
We therefore have from \eqref{35}
\begin{equation}\label{36}
  E(P)\le E_U(\al):=\frac{\al}{\al-1}E(Q)+\frac{1}{\al-1}\bar r(\al).
\end{equation}
The minimum of this upper bound
over all $\al$ in that range can be computed. Indeed, note that, as $\al\to1$
from the right, $E_U(\al)\to\iy$. Moreover, the derivative of $E_U$, that is given by
\begin{equation}
  E_U'(\al)=-\frac{E(Q)}{(\al-1)^2}-\frac{1}{(\al-1)^2}\Big[\frac12 a
  -\frac12\sqrt{a^2-4b^2p\al^2}\Big]+\frac1{\al-1}\frac{2b^2p\al}{\sqrt{a^2-4b^2p\al^2}},
\end{equation}
tends to $\iy$ as $\al\to a/(2b\sqrt p)$ from the left.
As a result, and since the equation
$E_U'(\al)=0$ turns out to have a unique root $\al^*$ in that range,
the minimizing $\al$ must be equal to $\al^*$. This unique root is given by
\begin{equation}
  \al^*=\frac{a^2\gamma_2+a\gamma_1\sqrt{\gamma_1^2+\gamma_2^2-a^2}}{\gamma_2(\gamma_1^2
  +\gamma_2^2)},
\end{equation}
where
\begin{equation}
  \gamma_1=a+2E(Q),
  \quad \gamma_2=2b\sqrt p.
\end{equation}
With this notation, the optimal upper bound of the form \eqref{36} is given by
\begin{equation}
  E_U=E_U(\al^*)=\frac{2\al^*E(Q)+a-\sqrt{a^2-\gamma_2^2(\al^*)^2}}{2(\al^*-1)}.
\end{equation}
As $b\to0$, we have $\al^*\to\iy$ and as a consequence $E_U\to E(Q)$.
That is, we recover the exponent $E(Q)$ as the fading intensity tends to zero.

As for a corresponding lower bound, we have
\begin{equation}
  E(P)\ge E_L(\al):=\frac{\al-1}{\al}E(Q)-\frac{1}{\al}\bar r(\al).
\end{equation}
A calculation shows that the maximizing $\al$ is
\begin{equation}
  \hat\al=\sqrt{\Big(\frac{a}{\gamma_2}\Big)^2-\Big(\frac{a^2}{\gamma_1\gamma_2}\Big)^2}
\end{equation}
and so
\begin{equation}
  E_L=E_L(\hat\al)=\frac{2\gamma_1(\hat\al-1)E(Q)-a\gamma_1+a^2}{2\gamma_1\hat\al}.
\end{equation}
As $b\to0$ we have $\hat\al\to\iy$ and so $E_L\to E(Q)$.

\subsection{Binary channel with erasure}\label{sec45}

We next consider the channel
\begin{equation}
Y_t=(a_tX_t)\oplus N_t,
\end{equation}
where $N_t$ is i.i.d.\  noise whereas $a_t$ is an erasure process.
Here, $a_t$, $X_t$ and $N_t$ take values in $\{0,1\}$ and $\oplus$ denotes
addition modulo $2$. It is assumed that
$\{a_t\}$ and $\{N_t\}$ are mutually independent.
We let $p=P(N_1=1)$ and assume $p\le1/2$.
The first model we examine
for $\{a_t\}$ is a hidden Markov model (an additional model appears
afterwards). Specifically,
we let $\{A_t\}$ be a stationary Markov process on the state space
$\{1,\ldots,d\}$ (independent of $(\{X_t\},\{N_t\})$) with a given transition
probability matrix $\PI$, assumed to be irreducible.
For a given function $f\{1,\ldots,d\}\to\{0,1\}$,
$a$ is given by $a_t=f(A_t)$, $t=1,\ldots,n$.
Denote by $P$ the probability measure induced by the above processes.
Let $Q$ denote a reference probability measure, under which
\begin{equation}
Y_t=X_t\oplus \tilde N_t,
\end{equation}
where, for each $n$, the law of the triplet $(\bX, \ba, \tilde \bN)$
is the same as that of $(\bX,\ba,\bN)$ under $P$
(in particular, the three are mutually independent under $Q$).

To calculate the R\'enyi divergence, note that
\begin{equation}
  P(\bX,\bY,\bA)=P(\bY|\bX,\bA)P(\bX)P(\bA)
  =\Big[\prod_{t=1}^nP(Y_t|a_tX_t)\Big]P(\bX)P(\bA),
\end{equation}
where for $(x,y)\in\{0,1\}^2$,
$P(y|x)=p$ if $y\ne x$ and $P(y|x)=1-p$ if $y=x$.
Also,
\begin{equation}
  Q(\bX,\bY,\bA)=\Big[\prod_{t=1}^nP(Y_t|X_t)\Big]P(\bX)P(\bA).
\end{equation}
Denoting by $P_n$ and $Q_n$ the respective laws of $(\bX,\bY,\bA)$,
we have
\begin{align}
  \frac{\al}{n}D_\al(Q_n\|P_n) &=
  \frac{1}{n(\al-1)}
  \ln \bE_{P}\Big[\Big(\prod_{t=1}^n\frac{P(Y_t|X_t)}{P(Y_t|a_tX_t)}\Big)^\al\Big]
  \\&=\frac{1}{n(\al-1)}
  \ln \bE_{P}\Big[\prod_{t:a_t=0,X_t=1}\Big(\frac{P(Y_t|X_t)}{P(Y_t|a_tX_t)}\Big)^\al
  \Big]\\
  &=\frac{1}{n(\al-1)}
  \ln \bE_{P}\prod_{t:a_t=0,X_t=1}
  \Big[p \Big(\frac{1-p}{p}\Big)^\al+(1-p)\Big(\frac{p}{1-p}\Big)^\al
  \Big]\\
  &\le\frac{1}{n(\al-1)}
  \ln \bE_{P}\prod_{t:a_t=0}\del(\al)\\
  &=\frac{1}{n(\al-1)}
  \ln \bE_{P}\Big[\del(\al)^{n-\sum_{t=1}^n a_t}\Big],
\end{align}
where we use the fact that $\del(\al):=
p (\frac{1-p}{p})^\al+(1-p)(\frac{p}{1-p})^\al\ge1$.
Let $\bar f(i)=1-f(i)$, $i=1,\ldots,d$, and denote by
\begin{equation}
Z_n=\frac{1}{n}\sum_{t=1}^n1_{\{a_t=0\}}=\frac{1}{n}\sum_{t=1}^n\bar f(A_t)
\end{equation}
the frequency of times $t$ when $a_t=0$.
Then we can write the above as
\begin{align}\label{50}
  \frac{\al}{n}D_\al(Q_n\|P_n) \le
  \frac{1}{n(\al-1)}
  \ln \bE_{P}[e^{n Z_n\ln\del(\al)}].
\end{align}
By similar considerations one obtains
\begin{equation}\label{51}
  \frac{\al-1}{n}D_\al(P_n\|Q_n)\le\frac{1}{n\al}\ln\bE_P[e^{n Z_n\ln\del(\al)}].
\end{equation}

For $\la\in\R$, let $\PI_\la=(\pi_\la(i,j))$, where
\begin{equation}
  \pi_\la(i,j)=\pi(i,j)e^{\la\bar f(j)},\qquad i,j\in\{1,\ldots,d\}.
\end{equation}
Then $\PI_\la$ is an irreducible matrix for every $\la$ and, by the
Perron-Frobenius theorem, has a real positive eigenvalue, denoted by $\rho(\PI_\la)$,
that dominates all eigenvalues in absolute value.
It is known that the random variables $Z_n$ satisfy the large deviation principle
with the good rate function $I:\R\to[0,\iy]$, defined as
\begin{equation}
  I(x)=\sup_{\la\in\R}\{\la x-\ln\rho(\PI_\la)\}
\end{equation}
(for the terminology see \cite{DZ}; for the above result see Theorem 3.1.2 therein).
Thus by Varadhan's lemma (Theorem 4.3.1 of \cite{DZ}), it follows that
\begin{equation}
  \lim_{n\to\iy}\frac{1}{n}\ln \bE_{P}[ e^{n Z_n\ln\del(\al)}]
  =\sup_{x\in\R}[x\ln\del(\al)-I(x)].
\end{equation}
We thus have
\begin{theorem}
  \label{th44}
For $Q$ the binary channel and $P$ the binary channel with erasure described
above, for every $\al>1$,
\begin{equation}\label{26}
  E(P)\le\frac{\al}{\al-1}E(Q)+\frac{1}{\al-1}\sup_{x\in\R}[x\ln\del(\al)-I(x)],
\end{equation}
and
\begin{equation}\label{27}
  E(P)\ge\frac{\al-1}{\al}E(Q)-\frac{1}{\al}\sup_{x\in\R}[x\ln\del(\al)-I(x)].
\end{equation}
\end{theorem}

\subsubsection*{Bounded fraction of erasures}
We now examine another model for the erasure
process $\{a_t\}$. In this model, the erasure process satisfies a single hard constraint,
namely that the relative number of erasures
$Z_n=\frac{1}{n}\sum_{t=1}^n1_{\{a_t=0\}}$ is a.s.-bounded. Specifically,
for some constant $z\in[0,1]$, it is assumed that $Z_n\le z$ a.s., for every $n$.
To relate this to the previous model,
note that this may occur when the (stationary, Markov)
process $A_t$ taking values in $\{1,\ldots,d\}$ is cyclic,
and where the subset $S\subset\{1,\ldots,d\}$
of states corresponding to erasure has cardinality $k$ with $k/d\le z$.
Of course, the class of processes $a$ satisfying the current
assumption is much broader.

Note that \eqref{50} and \eqref{51} are valid.
As a result, we obtain in this case, for $\al>1$,
\begin{equation}
  \frac{\al-1}{\al}E(Q)-\frac{1}{\al}z\ln\del(\al)\le
  E(P)\le \frac{\al}{\al-1}E(Q)+\frac{1}{\al-1}z\ln\del(\al).
\end{equation}
Clearly, this model has a property analogous to that established for the channel
with fading, namely that as $z\to0$, both bounds converge (upon optimization
with respect to $\al$) to $E(Q)$.

\section{Other applications}\label{sec5}

\subsection{Rate--distortion coding}\label{sec51}

Consider the problem of rate--distortion coding of a source sequence
$Y_1,Y_2,\ldots$ given by
\begin{equation}
Y_t=X_t+Z_t,
\end{equation}
where, under the probability measure $P$, $\{X_t\}$ is an i.i.d., $\calN(0,\sigma^2)$
process and $\{Z_t\}$ is a process that is independent of $\{X_t\}$.
For simplicity, assume the random vector $\bZ=(Z_1,\ldots,Z_n)$
has density, denoted $f_Z$.
Each source sequence $\by=(y_1,\ldots,y_n)$ is compressed to a string of
$nR$ nats, from which the decoder reconstructs an approximated sequence
$\hat{\by}=(\hat{y}_1,\ldots,\hat{y}_n)$.
We are interested in a lower bound on
\begin{equation}
P(\calE_{n,d}),\qquad \calE_{n,d}:=\Big\{\sum_{t=1}^n(Y_t-\hat{Y}_t)^2>nd\Big\},
\end{equation}
where $d$ is large enough so that this probability decays exponentially.

The joint density of $(\bY,\bZ)$ under $P$ is thus given by
$g(\by-\bz)f_Z(\bz)$, where $g$ is the i.i.d.\ $\calN(0,\sigma^2)$)
density. We consider a reference measure $Q$, under which the joint density
of $(\bY,\bZ)$ is $g(\by)f_Z(\bz)$.
Since the event $\calE_{n,d}$ is measurable on the sigma-field of $\bY$,
and under $Q$, $\bY$ and $\bZ$ are mutually, independent,
the law of $\bZ$ is irrelevant for the estimation of $Q(\calE_{n,d})$,
in the sense that $Q(\calE_{n,d})=G(\calE_{n,d})$, where we denote by $G$
the law of $\bY$ under $Q$ (equivalently, that of $\bX$ under $P$).
In the appendix, we show that
\begin{equation}
\label{app}
\liminf_{n\to\infty}\frac {\ln G(\calE_{n,d})}{n}\ge -\Ph[R-R_G(d)],
\end{equation}
where $R_G(d)=\frac{1}{2}\ln\frac{\sigma^2}{d}$ is the rate--distortion
function of the Gaussian source $\{X_t\}$ and
\begin{equation}
\Ph(u)\dfn\frac{e^{2u}-1}{2}-u.
\end{equation}

We now calculate the divergence term. With $P_n$ and $Q_n$ denoting the respective
laws of $(\bY,\bZ)$,
\begin{eqnarray}\label{67}
\frac{\alpha}{n}D_\alpha(Q_n\|P_n)&=&\frac{1}{n(\al-1)}\ln\left[
\int_{\reals^n}\mbox{d}\bz f_Z(\bz)\int_{\reals^n}\mbox{d}\by\cdot
g^{\alpha}(\by)g^{1-\alpha}(\by-\bz)\right]\nonumber\\
&=&\frac{1}{n(\al-1)}\ln\left[\int_{\reals^n}\mbox{d}\bz
f_Z(\bz)\exp\left\{\frac{\al(\al-1)\|\bz\|^2}{2\sigma^2}\right\}\right],
\end{eqnarray}
where the second step follows by appealing to identity \eqref{identity}.
The usefulness of the bound will now depend on estimating
the last expression. Obviously, for this expression to be finite,
the tails of $f_Z$ must decay faster than those of a Gaussian.

Consider the, for example, the case where $\sum_{t=1}^nZ_t^2\le nA^2$ almost surely.
In this case, the right-hand side
of \eqref{67} is bounded by $\frac{\al A^2}{2\sigma^2}$.
Using this bound together with \eqref{app} in \eqref{52} gives
\begin{eqnarray}
E(R,P)\le \inf_{\al\ge 1}\left[\frac{\al \Ph[R-R_G(d)]}{\al-1}+\frac{\al
A^2}{2\sigma^2}\right]=\left(\sqrt{\Ph[R-R_G(d)]}+\frac{A}{\sqrt{2}\sigma}\right)^2.
\end{eqnarray}
In a similar way, one obtains
\begin{equation}
E(R,P)\ge \left(\sqrt{\Ph[R-R_G(d)]}-\frac{A}{\sqrt{2}\sigma}\right)^2.
\end{equation}

An analogous derivation can be made for the case where $\{X_t\}$ is a
binary memoryless source with parameter $p$,
$\{Z_t\}$ is a binary interference with
normalized Hamming weight limited by $A$, and
$Y_t=X_t\oplus Z_t$. We then end up with
\begin{equation}
E(R)\le \inf_{\al>1}\left[\frac{\al F(R,D)}{\al-1}+\frac
{A\ln[p^{\al}(1-p)^{1-\al}+(1-p)^{\al}p^{1-\al}]}{\al-1}\right],
\end{equation}
where $F(R,D)$ is the source coding error exponent \cite{Marton74}
associated with $\{X_t\}$.

\subsection{Extension to a pair of sources}

A possible extension of this example is associated with the problem of
separate encodings and joint decoding of correlated sources. Let
$\{(X_i,Y_i)\}_{i=1}^n$ be $n$ independent copies of a random pair
$(X,Y)$ distributed according to $P_{XY}(x,y)$, $x\in\calX$, $y\in\calY$.
The sequences $\{X_i\}$ and $\{Y_i\}$ are compressed separately by two
encoders (that do not cooperate) at rates $R_x$ and $R_y$,
respectively.
The respective compressed bit--streams are both fed into a joint decoder
that produces reconstructions $\{\hat{X}_i\}$ and $\{\hat{Y}_i\}$,
whose components take on values in alphabets $\hat{\calX}$ and
$\hat{\calY}$, respectively.
Let $\rho_x:\calX\times\hat{\calX}\to\reals$ and
$\rho_y:\calY\times\hat{\calY}\to\reals$ be given distortion functions.
We are interested in a lower bound on
\begin{equation}
P\left\{\sum_{i=1}^n\rho_x(X_i,\hat{X}_i)\ge nd_x,~
\sum_{i=1}^n\rho_y(Y_i,\hat{Y}_i)\ge nd_y\right\}
\end{equation}
for some prescribed distortion levels $d_x$ and $d_y$.
We wish to pass to a reference source for which
$\{X_i\}$ and $\{Y_i\}$ are statistically independent, that is,
$Q_{XY}(x,y)=Q_X(x)Q_Y(y)$.
Under $Q$, the probability of the above event decays exponentially
at rate $F_x^Q(R_x,d_x)+F_y^Q(R_y,d_y)$, where $F_x^Q$ and $F_y^Q$ are the
source coding exponents of the separate reference sources,
$Q_X$ and $Q_Y$, respectively.
Thus, our upper bound on the exponent is given by
\begin{align}\notag
&E(R_x,R_y,d_x,d_y)\\
&\qquad\le\inf_{\al>1}\inf_{Q_X,Q_Y}\left\{\frac{\al}{\al-1}
[F_x^Q(R_x,d_x)+F_y^Q(R_y,d_y)]+\al D_{\al}(Q_X\times Q_Y\|P_{XY})\right\}.
\end{align}
In this setting, to the best of our knowledge,
there does not exist any competing bound in the literature.

\subsection{The problem of guessing}

Let $\bY=(Y_1,\ldots,Y_n)$ be a random vector with a given distribution.
Let $\hat{\bY}_1,\hat{\bY}_2,\ldots$ be a sequence of `guesses' of the random vector
$\bY$ that is generated without observing $\bY$.
within distortion $d$ from $\by$,
Denoting by $\rho$ the Hamming distance and fixing a distortion level $d\ge0$,
let $\Gam(\bY)$ denote the number of trials it takes to correctly guess $\bY$
within distortion level $d$, i.e.,
\begin{equation}
\Gam(\bY)=\min\{i:~\rho(\bY,\hat{\bY}_i)\le nd\}.
\end{equation}
In \cite{arimer}, it was shown that for a given
discrete memoryless source $Q$ and a given parameter $\la > 0$,
\begin{equation}\label{68}
\liminf_{n\to\iy}\frac{1}{n}\ln\bE_Q\{\Gam(\bY)^\la\}\ge
\sup_{\hat Q_1}[\la R(d,\hat Q_1)-D(\hat Q_1\|Q_1)],
\end{equation}
where $Q_1$ denotes the marginal of $Q$, $R(\cdot,\hat Q_1)$ denotes the
rate--distortion function of the source $\hat Q_1$,
and the supremum is over $\hat Q$
in the set of probability measures over the alphabet of $Y_1$.

Using the comparison bounds, we can estimate this quantity for a more general model.
Specifically, consider the model discussed at the end of Subsection \ref{sec51}.
Namely, $Y_t$ is binary and takes the form $Y_t=X_t\oplus Z_t$,
where, under a probability measure $P$, $\{X_t\}$ and $\{Z_t\}$ are mutually independent,
and $X_t$ are i.i.d.\  with parameter $p$. Assuming that the normalized number of times
$t$ when $Z_t=1$ is bounded by a constant $A$, the R\'enyi divergence term is bounded
by
\begin{equation}\label{69}
  \frac{\al}{n}D_\al(Q_n\|P_n)\le
  \frac{A}{\al-1}\ln[p^{\al}(1-p)^{1-\al}+(1-p)^{\al}p^{1-\al}],
\end{equation}
where as before, $P_n$ and $Q_n$ are the respective laws of $(\bY,\bZ)$.
We can now appeal to \eqref{31}. Using this inequality
(with the roles of $P$ and $Q$ interchanged), we have for arbitrary $\al>1$
and denoting $s=(\al-1)/\al$,
\begin{align}\label{70}
\frac{1}{n}\ln \bE_P\{\Gam(\bY)^\rho\}
\ge\frac{\al}{n(\al-1)}\ln\bE_Q\{\Gam(\bY)^{s\rho}\}-\frac{\al}{n}D_\al(Q_n\|P_n).
\end{align}
Using \eqref{68} and \eqref{69} in \eqref{70} gives
\begin{align}
&\liminf_{n\to\iy}\frac{1}{n}\ln \bE_P\{\Gam(\bY)^\rho\}\notag
\\ \notag
&\qquad\ge
\sup_{\al> 1}\Big\{\sup_{\hat Q_1}
\Big[\rho R(d,\hat Q_1)-\frac{\al}{\al-1}D(\hat Q_1\|Q_1)\Big]\\
& \qquad\qquad
-\frac{A}{\al-1}\ln[p^{\al}(1-p)^{1-\al}+(1-p)^{\al}p^{1-\al}]\Big\}.
\end{align}

\appendix

\section{Appendix}\beginsec
\manualnames{A}
\setcounter{equation}{0}

\subsection{Proof of the LPCB in a simple case}
Here we prove the LPCB is the case where the support is a finite set
(see \cite{ACD14} for the general setting).
Let $\calX$ be a finite set, let $\{P_i\}_{i\in\calX}$ and $\{Q_i\}_{i\in\calX}$
be two probability distributions defined on it and let $G:\calX\to[0,\iy)$ be a given
function.
\begin{proposition}\label{prop1}
Assume $\sum_{i\in\calX}G_iP_i>0$ and $\sum_{i\in\calX}G_iQ_i>0$.
Then for all $\al>1$
\begin{equation}\label{82}
  \frac{1}{\al-1}\ln \sum_i G_i^{\al-1}P_i\le\frac{1}{\al}\ln\sum_i G_i^\al Q_i
  +D_\al(P\|Q).
\end{equation}
Moreover, given $P$, $G$ and $\al$ as above, there exists $Q$ for which \eqref{82}
holds with equality.
\end{proposition}

\proof
When one does not have $P\ll Q$, the divergence term above equals $+\iy$ by definition,
and there is nothing to prove. Hence assume $P\ll Q$. Denote by $S_P$, $S_Q$
and $S_G$ the support of $P$, $Q$ and $G$, respectively. Let $S=S_Q\cap S_G$.
Using H\"older's inequality with the exponents $\al$ and $\al/(\al-1)$
and measure $\{Q_i\}_{i\in S}$,
\begin{align}
  \sum_SG_i^{\al-1}P_i
  &=\sum_S\frac{P_i}{Q_i}G_i^{\al-1}Q_i\\
  &\le\Big(\sum_S\Big(\frac{P_i}{Q_i}\Big)^{\al}Q_i\Big)^{1/\al}
  \Big(\sum_S\Big(G_i^{\al-1}\Big)^{\al/(\al-1)}Q_i\Big)^{(\al-1)/\al}\\
  &=\Big(\sum_S\Big(\frac{P_i}{Q_i}\Big)^{\al}Q_i\Big)^{1/\al}
  \Big(\sum_SG_i^\al Q_i\Big)^{(\al-1)/\al}.
\end{align}
Thus
\begin{align}
  \Big(\sum_SG_i^{\al-1}P_i\Big)^\al
  \Big(\sum_SG_i^\al Q_i\Big)^{(1-\al)}
  &\le\sum_S\Big(\frac{P_i}{Q_i}\Big)^{\al}Q_i\\
  &\le\sum_{S_Q}\Big(\frac{P_i}{Q_i}\Big)^{\al}Q_i.
\end{align}
For $i$ not in $S$, $G_i^\al Q_i=0$, and because $P\ll Q$, also
$G_i^{\al-1}P_i=0$. Thus,
on the left-hand side, the summation can be performed over all of $\calX$.
As a result, taking logarithms and dividing by $\al(\al-1)$,
using the definition of the divergence \eqref{12} gives the inequality \eqref{82}.
To show the final assertion set
$Q_i=G_i^{-1}P_i/Z$ for $i\in S_P\cap S_G$ and $0$ off of that set. Here,
$Z=\sum_{i\in S_P\cap S_G}G_i^{-1}P_i>0$ by assumption. Substituting in
\eqref{82} gives equality by a direct calculation.
\qed

\subsection{Proof of Theorem \ref{th31}}
A bound on the divergence between any two univariate Gaussians is deduced from identity
\eqref{identity} as follows. Given $x\in\R$, $\xi\in\R$ such that $|\xi|\le \Gam$,
and any $\al>1$ and $s>1-1/\al$,
\begin{align}
&D_\al(\calN(x,\sigma^2/s)\|\calN(x+\xi,\sigma^2))\notag\\
&\quad=\frac{1}{\al(\al-1)}\ln
\left[\frac{s^{\al/2}}{\sqrt{1+\al(s-1)}}\cdot
\exp\left\{-\frac{\al(1-\al)s\xi^2}{2\sigma^2[1+\al(s-1)]}\right\}\right]\nonumber\\
&\quad=\frac{1}{\al(\al-1)}\left\{\frac{\al\ln
s}{2}-\frac{\ln[1+\al(s-1)]}{2}+\frac{\al(\al-1)s\xi^2}{2\sigma^2[1+\al(s-1)]}\right\}\nonumber\\
&\quad=\frac{\ln s}{2(\al-1)}-
\frac{\ln[1+\al(s-1)]}{2\al(\al-1)}+\frac{s\xi^2}{2\sigma^2[1+\al(s-1)]}\nonumber\\
&\quad\le\frac{\ln s}{2(\al-1)}-
\frac{\ln[1+\al(s-1)]}{2\al(\al-1)}+\frac{s\Gam^2}{2\sigma^2[1+\al(s-1)]}.
\end{align}

Let $P_n$ and $Q_n$ denote the respective probability laws of $(\bX,\bY)$. Then
\begin{eqnarray}
D_\alpha(Q_n\|P_n)&=&
\frac{1}{\al(\al-1)}\ln\bE_P\Big[\Big(\frac{Q(\bX,\bY)}{P(\bX,\bY)}\Big)^\al\Big]
\\
&=&
\frac{1}{\al(\al-1)}\ln\sum_{\bx}\int \dd\by
\Big(\frac{Q(\by|\bx)}{P(\by|\bx)}\Big)^\al P(\by|\bx)\Pi(\bx)
\\
&=&\frac{1}{\al(\al-1)}\ln\sum_{\bx}\Pi(\bx)\left[\int_{\reals^n}\mbox{d}\by
(2\pi\sigma^2/s)^{-\al
n/2}(2\pi\sigma^2)^{-(1-\al)n/2}\times\right.\nonumber\\
& &\left.\exp\left\{-\frac{s\al}{2\sigma^2}\sum_{t=1}^n(y_t-x_t)^2\right\}\cdot
\exp\left\{-\frac{1-\al}{2\sigma^2}\sum_{t=1}^n[y_t-x_t-g_t(x^n,y^{t-1})]^2\right\}\right]\nonumber\\
&=&\frac{n\ln s}{2(\al-1)}-\frac{n\ln(2\pi\sigma^2)}{2\al(\al-1)}+
\frac{1}{\al(\al-1)}\ln\sum_{\bx}\Pi(\bx)\left[\int_{\reals^n}\mbox{d}\by\times\right.\nonumber\\
& &\left.\exp\left\{-\sum_{t=1}^n\left(\frac{s\al}{2\sigma^2}[y_t-x_t]^2
+\frac{1-\al}{2\sigma^2}[y_t-x_t-g_t(x^n,y^{t-1})]^2\right)\right\}\right]\\
&\dfn&\frac{n\ln s}{2(\al-1)}-\frac{n\ln(2\pi\sigma^2)}{2\al(\al-1)}+
\frac{1}{\al(\al-1)}\cdot Z_n
\end{eqnarray}
Let us focus on the expression of $Z_n$.
For $t\in\{1,\ldots,n\}$ let $\Pi(\xi^t)=\sum_{x^n:x^t=\xi^t}\Pi(x^n)$.
Then
\begin{eqnarray}
Z_n&=&\ln\sum_{\bx}\Pi(\bx)\left[\int_{\reals^{n-1}}\mbox{d}y^{n-1}\times\right.\nonumber\\
& &\left.\exp\left\{-\sum_{t=1}^{n-1}\left(\frac{s\al}{2\sigma^2}[y_t-x_t]^2
+\frac{1-\al}{2\sigma^2}[y_t-x_t-g_t(x^n,y^{t-1})]^2\right)\right\}\times\right.\nonumber\\
& &\left.\int_\reals\mbox{d}y_n
\exp\left\{-\left(\frac{s\al}{2\sigma^2}(y_n-x_n)^2
+\frac{1-\al}{2\sigma^2}[y_n-x_n-g_n(x^n,y^{n-1})]^2\right)\right\}\right]\\
&=&\ln\sum_{\bx}\Pi(\bx)\left[\int_{\reals^{n-1}}\mbox{d}y^{n-1}\times\right.\nonumber\\
& &\left.\exp\left\{-\sum_{t=1}^{n-1}\left(\frac{s\al}{2\sigma^2}[y_t-x_t]^2
+\frac{1-\al}{2\sigma^2}[y_t-x_t-g_t(x^n,y^{t-1})]^2\right)\right\}\times\right.\nonumber\\
& &\left.\sqrt{\frac{2\pi\sigma^2}{1+\al(s-1)}}
\exp\left\{\frac{s\al(\al-1)g_n^2(x^n,y^{n-1})}{2\sigma^2[1+\al(s-1)]}\right\}\right]\\
&\le&\ln\sum_{x^{n-1}}\Pi(x^{n-1})\left[\int_{\reals^{n-1}}\mbox{d}y^{n-1}\times\right.\nonumber\\
& &\left.\exp\left\{-\sum_{t=1}^{n-1}\left(\frac{s\al}{2\sigma^2}[y_t-x_t]^2
+\frac{1-\al}{2\sigma^2}[y_t-x_t-g_t(x^n,y^{t-1})]^2\right)\right\}\right]+\nonumber\\
& &\frac{1}{2}\ln\left[\frac{2\pi\sigma^2}{1+\al(s-1)}\right]+
\frac{s\al(\al-1)\max_{x^n,y^{n-1}}g_n^2(x^n,y^{n-1})}{2\sigma^2[1+\al(s-1)]}\\
&\le&Z_{n-1}+\frac{1}{2}\ln\left[\frac{2\pi\sigma^2}{1+\al(s-1)}\right]
+\frac{s\al(\al-1)\Gam_n^2}{2\sigma^2[1+\al(s-1)]}.
\end{eqnarray}
 From this recursion on $Z_n$, we have
\begin{eqnarray}
Z_n&\le&\frac{n}{2}\ln\left[\frac{2\pi\sigma^2}{1+\al(s-1)}\right]
+\frac{s\al(\al-1)\sum_{t=1}^n\Gam_t^2}{2\sigma^2[1+\al(s-1)]}\\
&\le&\frac{n}{2}\ln\left[\frac{2\pi\sigma^2}{1+\al(s-1)}\right]
+\frac{ns\al(\al-1)\Gam^2}{2\sigma^2[1+\al(s-1)]}.
\end{eqnarray}
Therefore
\begin{align}
&D_\alpha(Q_n\|P_n)\\
&=
\frac{n\ln s}{2(\al-1)}-\frac{n\ln(2\pi\sigma^2)}{2\al(\al-1)}+
\frac{1}{\al(\al-1)}\cdot Z_n\nonumber\\
&\le\frac{n\ln s}{2(\al-1)}-\frac{n\ln(2\pi\sigma^2)}{2\al(\al-1)}+
\frac{1}{\al(\al-1)}\left\{\frac{n}{2}\ln\left[\frac{2\pi\sigma^2}{1+\al(s-1)}\right]
+\frac{ns\al(\al-1)\Gam^2}{2\sigma^2[1+\al(s-1)]}\right\}\nonumber\\
&=\frac{n\ln s}{2(\al-1)}-
\frac{n\ln[1+\al(s-1)]}{2\al(\al-1)}
+\frac{ns\Gam^2}{2\sigma^2[1+\al(s-1)]}.
\label{57}
\end{align}
Substituting in \eqref{45}, using the bound $E(R,Q_s,d)\le E_{\mbox{\tiny sl}}(R,Q_s)$
for every $d$, and finally optimizing over $s$ and $\al$, yields \eqref{49}.
\qed

\subsection{Proof of Theorem \ref{th45}}

As a reference, we will use the models $Q=Q_{\phi,\theta}$, under which
\begin{equation}
  Y_t=\phi X_t+\tilde W_t,
\end{equation}
where $\{\tilde W_t\}$ are i.i.d.\  $\calN(0,\theta)$, independent of $\{X_t\}$.
Here, $\phi>0$ and $\theta>0$ are parameters.
Note that, for each of the models $Q$, $d$ is the optimal decoding metric.
One has
\begin{equation}
P(\by|\bx)=(2\pi\sigma^2)^{-n/2}\prod_{t=1}^n
\exp\left\{-\frac{1}{2\sigma^2}\left(y_t-x_t-\sum_{i=1}^kh_ix_{t-1}\right)^2\right\},
\end{equation}
and $Q(\by|\bx)=\prod_{t=1}^nQ(y_t|x_t)$, where $Q(y|x)$ is given by
\begin{equation}
Q(y|x)=\sqrt{\frac{\theta}{\pi}}\cdot\exp\{-\theta(y-\phi x)^2\},~~~\theta> 0,~\phi> 0.
\end{equation}
In order to calculate the R\'enyi divergence, we use the identity \eqref{identity}
with the assignments: $a=\alpha/(2\sigma^2)$, $b=(1-\al)\theta$,
$u=x_t+\sum_{i=1}^kh_kx_{t-i}$ and $v=\phi x_t$, to get, under the assumption
\begin{equation}
  a+b=\frac{\al}{2\sigma^2}+(1-\alpha)\theta>0,
\end{equation}
\begin{eqnarray}
& &\int_{\reals^n}\mbox{d}\by\cdot\prod_{t=1}^n\exp\left\{
-\frac{\alpha}{2\sigma^2}\left(y_t-x_t-\sum_{i=1}^kh_ix_{t-i}\right)^2
-(1-\al)\theta(y_t-\phi x_t)^2
\right\}\\
&=&\left[\frac{\pi}{(1-\al)\theta+\al/2\sigma^2}\right]^{n/2}
\cdot\exp\left\{-\frac{\al(1-\al)\theta
\sum_t\left[(1-\phi)x_t+\sum_{i=1}^kh_ix_{t-i}\right]^2}
{\al+2(1-\al)\theta\sigma^2}\right\}\\
&=&\left[\frac{2\pi\sigma^2}{\al+2(1-\al)\theta\sigma^2}\right]^{n/2}
\cdot\exp\left\{-\frac{n\al(1-\al)\theta S
[(1-\phi)^2+2(1-\phi)r_1+r_2]}{\al+2(1-\al)\theta\sigma^2}\right\}.
\end{eqnarray}
Therefore
\begin{align}
& \frac{1}{n}D_\al(P_n\|Q_n)\nonumber\\
&=\frac{1}{n\al(\al-1)}\ln\left[\left(\frac{\theta}{\pi}\right)^{n(1-\al)/2}
(2\pi\sigma^2)^{-n\al/2}
\left(\frac{2\pi\sigma^2}{\al+2(1-\al)\theta\sigma^2}\right)^{n/2}
\right.\nonumber\\
& \qquad\left.\times\exp\left\{-\frac{n\al(1-\al)\theta S
[(1-\phi)^2+2(1-\phi)r_1+r_2]}{\al+2(1-\al)\theta\sigma^2}\right\}\right]\nonumber\\
&=\frac{1}{n\al(\al-1)}\ln\left[
\frac{(2\theta\sigma^2)^{n(1-\alpha)/2}}{(\al+2(1-\al)\theta\sigma^2)^{n/2}}
\cdot\exp\left\{-\frac{n\al(1-\al)\theta S
[(1-\phi)^2+2(1-\phi)r_1+r_2]}{\al+2(1-\al)\theta\sigma^2}\right\}\right]\nonumber\\
&=-\frac{\ln(2\theta\sigma^2)}{2\al}-\frac{\ln(\al+2(1-\al)\theta\sigma^2)}{2\al(\al-1)}
+\frac{\theta S[(1-\phi)^2+2(1-\phi)r_1+r_2]}{\al+2(1-\al)\theta\sigma^2}.
\end{align}
For a code of rate zero operating over the reference channel $Q$,
the best achievable exponent is known to be
\begin{align}
E(0,Q,d)=\frac{S\theta\phi^2}{2},
\end{align}
where we have used an extension of the zero--rate lower bound of
\cite{SGB1}, \cite{SGB2}
that applies to codes with a given composition $\mu$ (see Sections 2 and 4 of \cite{mer1}).
Then we have a lower bound from \eqref{45}, for $\al>1$,
\begin{equation}
  E(0,P,d)\ge\frac{\al-1}{\al}E(0,Q,d)-(\al-1)\Del^{P,Q}_\al.
\end{equation}
Thus
\begin{eqnarray}
E(0,P,d)&\ge&\sup_{(\alpha,\theta,\phi)\in\calS}\left[
\frac{(\al-1)S\theta\phi^2}{2\al}
+\frac{(\al-1)\ln(2\theta\sigma^2)+\ln(\al+2(1-\al)\theta\sigma^2)}
{2\al}\right.\nonumber\\
& &\qquad\qquad\left.-\frac{(\al-1)\theta S[(1-\phi)^2+2(1-\phi)r_1+r_2]}{\al+2(1-\al)\theta\sigma^2}\right],
\end{eqnarray}
where
$$\calS=\left\{(\al,\theta,\phi):~\al> 1,~\theta<\frac{\al}{2(\al-1)\sigma^2},
~\phi>0\right\}.$$
The maximization over $\phi$ is simple since the objective is quadratic in
$\phi$. In particular, the part that depends on $\phi$ is of the form
$A\phi^2+B\phi$, where
\begin{equation}
A=-\frac{\theta S(\al-1)(\al+2(\al-1)\theta\sigma^2)}{2\al(\al+2(1-\al)\theta\sigma^2)}<0,
\end{equation}
and
\begin{equation}
B=\frac{2\theta S(\al-1)(1+r_1)}{\al+2(1-\al)\theta\sigma^2}.
\end{equation}
The maximum of $A\phi^2+B\phi$ is
\begin{equation}
-\frac{B^2}{4A}=-\frac{2\theta \al(\al-1)S(1+r_1)^2}
{4(\al-1)^2\theta^2\sigma^4-\al^2},
\end{equation}
and our lower bound becomes,
\begin{align}
\notag
&\frac{(\al-1)\ln(2\theta\sigma^2)+\ln(\al+2(1-\al)\theta\sigma^2)}{2\al}
-\frac{(\al-1)\theta S[1+2r_1+r_2]}{\al+2(1-\al)\theta\sigma^2}
\\
&\qquad -\frac{2\theta \al(\al-1)S(1+r_1)^2}
{4(\al-1)^2\theta^2\sigma^4-\al^2}.
\end{align}
It would be more convenient to define $\theta=\tau\al/[2(\al-1)\sigma^2]$,
$\tau\in(0,1)$, and to transform the parameter set from $(\al,\theta)$ to
$(\al,\tau)$. Denoting
\begin{equation}
\Ph(\al,\tau):=\frac{(\al-1)\ln[\tau\al/(\al-1)]+\ln[\al(1-\tau)]}{2\al},
\end{equation}
the expression is then
\begin{align}
\notag
&\Ph(\al,\tau)
+\frac{S\tau}{2\sigma^2}\Big[-\frac{1+2r_1+r_2}{1-\tau}+\frac{2(1+r_1)^2}{1-\tau^2}\Big]\\
&=\Ph(\al,\tau)+\frac{S}{2\sigma^2}\Big[-\frac{\tau}{1-\tau}(r_2-r_1^2)
+\frac{\tau}{1+\tau}(1+r_1)^2\Big],
\label{72}
\end{align}
to be maximized over $(\tau,\al)\in(0,1)\times(1,\iy)$.
Now the function $\Ph(\al,\tau)$
is always non--positive (the maximum over $\tau\in(0,1)$ for a given $\alpha$ is zero)
and it vanishes for $\alpha=1/(1-\tau)$ (hence this is the optimum choice
of $\alpha)$. Thus, we are left with maximizing the second term of \eqref{72} over $\tau$.
Recall that $r_2\ge r_1^2$, and that
$a=r_2-r_1^2\ge0$ and $b=(1+r_1)^2$. The maximum is given by
\begin{equation}
E(0,P)\ge
  \begin{cases}
    \frac{S}{4\sigma^2}(\sqrt b-\sqrt a)^2
    =\frac{S}{4\sigma^2}[|1+r_1|-\sqrt{r_2-r_1^2}]^2 & a\le b,
    \\
    0 & \text{otherwise.}
  \end{cases}
\end{equation}
This establishes the first inequality in \eqref{81}.

In the case $a<b$, the maximizing $\tau$ is given by
$(\sqrt b-\sqrt a)^2/(b-a)\in(0,1)$. If we use this in the expression for
the optimal $\al$ and $\theta$,
we obtain that the optimal $\theta$ is $1/(2\sigma^2)$ and the optimal $\phi^2$ is given
by $\phi^2=(\sqrt b+\sqrt a)^2$.
Thus under the selected reference model,
\begin{equation}
E(0,Q,d)=\frac{S\theta\phi^2}{2}=\frac{S}{4\sigma^2}(\sqrt b+\sqrt a)^2.
\end{equation}
By virtue of \eqref{59}, this gives
namely
\begin{equation}
  \sup_d\inf_{P\in F} E(0,P,d)\le\frac{S}{4\sigma^2}(\sqrt b+\sqrt a)^2.
\end{equation}
\qed

\subsection{Proof of Theorem \ref{th42}}
To work with the upper bound, we compute the R\'enyi divergence term,
\begin{equation}
\frac{\al}{n}D_\al(Q_n\|P_n)=\frac{1}{n(\al-1)}\ln \bE_{P}\Big[\Big(
\frac{Q(\bX,\bY,\btheta)}{P(\bX,\bY,\btheta)}\Big)^\al\Big].
\end{equation}
We have
\begin{equation}
P(\bX,\bY,\btheta)=P(\bY|\bX,\btheta)P(\bX)P(\btheta)
=\Big[\prod_{t=1}^ng(Y_t|(1+\theta_t)X_t)\Big]P(\bX)P(\btheta),
\end{equation}
where $g(y|x)=(2\pi\sigma^2)^{-1/2}e^{-(y-x)^2/(2\sigma^2)}$, and
\begin{equation}
Q(\bX,\bY,\btheta)=\Big[\prod_{t=1}^ng(Y_t|X_t)\Big]P(\bX)P(\btheta).
\end{equation}
Thus
\begin{align}\label{24}
\frac{\al}{n}D_\al(Q_n\|P_n)
&= \frac{1}{n(\al-1)}\ln \bE_{P}\Big[\Big(
\prod_{t=1}^n\frac{g(Y_t|X_t)}{g(Y_t|(1+\theta_t)X_t)}\Big)^\al\Big]\\
&= \frac{1}{n(\al-1)}
\ln \bE_{P}\Big[\exp\Big\{\frac{\al}{2\sigma^2}\sum_{t=1}^n[\theta_t^2X_t^2-2\theta_tX_t(Y_t-X_t)]
\Big\}\Big] \\
&= \frac{1}{n(\al-1)}
\ln \bE_{P}\Big[\exp\Big\{\frac{\al}{2\sigma^2}\sum_{t=1}^n[-\theta_t^2X_t^2
-2\theta_tX_tW_t]\Big\}\Big] \\
&= \frac{1}{n(\al-1)}
\ln \bE_{P}\Big[\exp\Big\{\frac{\al(\al-1)}{2\sigma^2}
\sum_{t=1}^n\theta_t^2X_t^2\Big\}\Big] \\
&\le \frac{1}{n(\al-1)}
\ln \bE_{P}\Big[\exp\Big\{\frac{\al(\al-1)\aaa^2}{2\sigma^2}
\sum_{t=1}^n\theta_t^2\Big\}\Big],
\label{24+}
\end{align}
where in the last line we have used the assumption $|X_t|\le\aaa$.

We have assumed that $\theta$ is a stationary, zero-mean Gaussian process.
Thus the limit
\begin{equation}
\lim_{n\to\iy}\frac{1}{n}\ln \bE_P\Big\{\exp\Big(c\sum_{t=1}^n\theta_t^2\Big)\Big\}
\end{equation}
can be computed using Szego's theorem (see \cite{gray}).
To this end, note first that the exponential moment is given by
\begin{equation}
\bE_P\Big\{\exp\Big(c\sum_{t=1}^n\theta_t^2\Big)\Big\}=\det(I-2cV_n)^{-1/2},
\end{equation}
where $V_{n}$ is the covariance matrix of $\theta_t$, $t=1,\ldots,n$.
Next, if $T_n$ is a sequence of Hermitian Toeplitz matrices of the form
$T_n=[t_{k-j};k,j=0,1,2,\ldots,n-1]$, where $t_k$ are absolutely summable,
and their spectral density
$f(\om)=\sum_{k=-\iy}^\iy t_ke^{-ik\om}$, $\om\in\R$,
satisfies $f(\om)\ge m>0$, $\om\in\R$, one has by Theorem 13 of \cite{gray}, that
\begin{equation}
\lim_{n\to\iy}\frac{1}{n}\ln\det(T_n)=\frac{1}{2\pi}\int_0^{2\pi}\ln f(\om)\dd\om.
\end{equation}
Recall that we assume that $r_k$ are absolutely summable.
Then, with $c=c(\al)=\al(\al-1)\aaa^2/(2\sigma^2)$, we obtain the bound
\begin{align}
\limsup_{n\to\iy}\frac{\al}{n}D_\al(Q_n\|P_n)&\le
\limsup_{n\to\iy}-\frac{1}{2(\al-1)n}\ln\det(I-2cV_n)\\
&=
-\frac{1}{4\pi(\al-1)}\int_0^{2\pi}\ln(1-2c\Sig_\theta(\om))\dd\om,
\end{align}
assuming $2c\sup_\om \Sig_\theta(\om)<1$.
As for the lower bound, a calculation similar to that of \eqref{24}--\eqref{24+} gives
\begin{align}
  \frac{\al-1}{n}D_\al(P_n\|Q_n)&= \frac{1}{n\al}\ln E_Q
  \Big[\exp\frac{\al(\al-1)}{2\sigma^2}\sum_{t=1}^n\theta_t^2X_t^2\Big]\\
  &\le \frac{1}{n\al}\ln E_Q
  \Big[\exp\frac{\al(\al-1)\aaa^2}{2\sigma^2}\sum_{t=1}^n\theta_t^2\Big].
\end{align}
Using the same considerations as before gives
\begin{equation}
  \limsup_{n\to\iy}\frac{\al-1}{n}D_\al(P_n\|Q_n)\le
  -\frac{1}{4\pi\al}\int_0^{2\pi}\ln(1-2c\Sig_\theta(\om))\dd\om,
\end{equation}
where again we assume that $\{\Sig_\theta\}$ satisfies $2c\sup_\om \Sig_\theta(\om)<1$.
\qed

\subsection{Proof of Theorem \ref{th43}}
Our estimates of the R\'enyi divergence are based on large deviation results
from \cite{bry-dem}. We first note that the divergence is given by
\begin{align}
  D_\al(Q_T\|P_T)&=\frac{1}{\al(\al-1)}\ln \bE_{P_T}[(Z_T)^{-\al}]\\
  &=\frac{1}{\al(\al-1)}\ln \bE_{P_T}
  \exp\Big[-\frac{\al}{\sigma}\int_0^t\theta_sX_s\dd W_s+\frac{\al}{2\sigma^2}
  \int_0^t(\theta_sX_s)^2\dd s\Big].
\end{align}
Note by \eqref{37} and \eqref{38} that
\begin{equation}
\sigma W_t=\int_0^t\theta_sX_s\dd s+\sigma\tilde W_t,
\end{equation}
and so
\begin{equation}
\int_0^t\theta_sX_s\dd W_s=\sigma^{-1}\int_0^t(\theta_sX_s)^2\dd s
+\int_0^t\theta_sX_s\dd\tilde W_s.
\end{equation}
Thus
\begin{align}
  D_\al(Q_T\|P_T)&=
  \frac{1}{\al(\al-1)}\ln \bE_{P_T}
  \exp\Big[-\frac{\al}{\sigma}\int_0^t\theta_sX_s\dd\tilde W_s-\frac{\al}{2\sigma^2}
  \int_0^t(\theta_sX_s)^2\dd s\Big].
\end{align}
Under $P_T$, conditioned on $(\theta_s,X_s, s\in[0,t])$, the integral
$\int_0^t\theta_sX_s\dd\tilde W_s$
is a Gaussian random variable with mean zero and variance $\int_0^t(\theta_sX_s)^2\dd s$.
Thus
\begin{align}
  D_\al(Q_T\|P_T)&=\frac{1}{\al(\al-1)}\ln \bE_{P_T}
  \exp\Big[\frac{\al(\al-1)}{2\sigma^2}\int_0^t(\theta_sX_s)^2\dd s\Big].
\end{align}
A similar calculation for
\begin{align}
  D_\al(P_T\|Q_T)&=\frac{1}{\al(\al-1)}\ln \bE_{Q_T}[(Z_T)^\al]\\
  &=\frac{1}{\al(\al-1)}\ln \bE_{Q_T}
  \exp\Big[\frac{\al}{\sigma}\int_0^t\theta_sX_s\dd W_s-\frac{\al}{2\sigma^2}
  \int_0^t(\theta_sX_s)^2\dd s\Big]
\end{align}
gives
\begin{align}
  D_\al(P_T\|Q_T)&=\frac{1}{\al(\al-1)}\ln \bE_{Q_T}
  \exp\Big[\frac{\al(\al-1)}{2\sigma^2}\int_0^t(\theta_sX_s)^2\dd s\Big].
\end{align}
As a result, the two divergences are equal, and
using $|X_t|\le\aaa$, we can bound them as follows:
\begin{align}
  D_\al(Q_T\|P_T)=D_\al(P_T\|Q_T)\le
  \frac{1}{\al(\al-1)}\ln
  \bE_{Q_T}\exp\Big[\frac{\al(\al-1)\aaa^2}{2\sigma^2}\int_0^T\theta_t^2\dd t\Big].
\end{align}
It is shown in Lemma 3 of \cite{bry-dem} that
\begin{equation}\label{29}
  \lim_{T\to\iy}\frac{1}{T}\ln \bE_{Q_T}\exp\Big[c\int_0^T\theta_t^2\dd t\Big]
  =-\frac{1}{4\pi}\int_{-\iy}^{\iy}\ln[1-4\pi c\Sig_\theta(\om)]\dd\om,
\end{equation}
provided that $c<1/(4\pi M)$. Specifically, \eqref{28} holds
since we have assumed that $p:=\aaa^2/(2\sigma^2)<1/(4\pi M)$.
Moreover, with $c(\al)=\al(\al-1)p$,
for any $\al>1$ such that $c(\al)<1/(4\pi M)$, we obtain from
\eqref{17} and \eqref{18} (by a derivation analogous to that of \eqref{52})
\begin{equation}\label{33}
  E(P)\le\frac{\al}{\al-1}E(Q)-\frac{1}{4\pi(\al-1)}\int_{-\iy}^\iy
  \ln[1-4\pi c(\al)\Sig_\theta(\om)]\dd\om
\end{equation}
\begin{equation}\label{34}
  E(P)\ge\frac{\al-1}{\al}E(Q)+\frac{1}{4\pi\al}\int_{-\iy}^\iy
  \ln[1-4\pi c(\al)\Sig_\theta(\om)]\dd\om.
\end{equation}
\qed

\subsection{Proof of \eqref{app}}

To prove (\ref{app}), we follow the main steps of \cite{Marton74}, with a little
twist since in our case the distortion measure (which is quadratic) is
unbounded. Consider an arbitrary rate--distortion code
$\calC=\{\hat{\by}_1,\ldots,\hat{\by}_M\}$, $M=e^{nR}$, $R$ being the coding
rate. Let us denote the event under discussion by
\begin{equation}
\calE=\left\{\by:~\min_m\sum_{t=1}^n(y_t-\hat{y}_{m,t})^2>nd\right\},
\end{equation}
where $\hat{y}_{m,t}$ is the $t$--th component of the reproduction
word $\hat{\by}_m$. Let
$R_G(d,\tilde{\sigma}^2)=\frac{1}{2}\ln\frac{\tilde{\sigma}^2}{d}$ denote the rate--distortion
function of the Gaussian memoryless source $\tilde{G}$ with variance $\tilde{\sigma}^2$.
We first show that under the assumption that $R_G(d,\tilde{\sigma}^2)> R$, there exists a
constant $\alpha(\tilde{\sigma}^2,d,R)> 0$ such that
$\tilde{G}(\calE)\ge\alpha(\tilde{\sigma}^2,d,R)$ for all sufficiently large $n$.
Let
\begin{equation}
\tilde{d}(\calC)\dfn
\frac{1}{n}\tilde{\bE}\{\min_m\|\bY-\hat{\bY}_m\|^2\},
\end{equation}
where $\tilde{\bE}$ denotes expectation under $\tilde{G}$. Let
$d_1=\tilde{\sigma}^2 e^{-2R}$ denote the optimum distortion of $\tilde{G}$ at
rate $R$. Then, obviously,
\begin{equation}
R_G(d,\tilde{\sigma}^2)>R=R_G(d_1,\tilde{\sigma}^2)\ge
R_G(\tilde{d}(\calC),\tilde{\sigma}^2),
\end{equation}
where the first inequality is by our assumption.
the equality is by definition of $d_1$ and the second inequality is due to the
fact that $\calC$ may not be optimal for $\tilde{G}$.
Since $R_G(\cdot,\tilde{\sigma}^2)$ is monotonically decreasing, then
\begin{equation}
d<d_1\le\tilde{d}(\calC).
\end{equation}
Now, let us denote $\delta(\by)=\min_m\|\by-\hat{\by}_m\|^2/n$
and let $d_0> d_1$ be an arbitrary large distortion level. Then,
assuming, without loss of generality, that the zero--vector belongs to
$\calC$, and so, $\delta(\by)\le\|\by\|^2/n$, we have:
\begin{eqnarray}
\tilde{d}(\calC)&\le&[1-\tilde{G}(\calE)]\cdot d+\int_{\by:~\delta(\by)\ge
d}\tilde{G}(\by)\delta(\by)\mbox{d}\by\nonumber\\
&=&[1-\tilde{G}(\calE)]\cdot d+\int_{\by:~d\le \delta(\by)\le
d_0}\tilde{G}(\by)\delta(\by)\mbox{d}\by+
+\int_{\by:~\delta(\by)\ge
d_0}\tilde{G}(\by)\delta(\by)\mbox{d}\by\nonumber\\
&\le&[1-\tilde{G}(\calE)]\cdot d+\tilde{G}(\calE)\cdot d_0+\frac{1}{n}\int_{\by:~\|\by\|^2\ge nd_0
}\tilde{G}(\by)\cdot\|\by\|^2\mbox{d}\by
\end{eqnarray}
Now, the last term, which is
\begin{equation}
\delta_n\dfn \frac{1}{n}\cdot(2\pi\tilde{\sigma}^2)^{-n/2}\int_{\|\by\|^2\ge
nd_0}\|\by\|^2\cdot\exp\{-\|\by\|^2/2\tilde{\sigma}^2\}\mbox{d}\by,
\end{equation}
is easily shown\footnote{Apply the
Chernoff bound and use the fact that
$\|\by\|^2e^{-s\|\by\|^2}$ is the negative derivative of
$e^{-s\|\by\|^2}$ w.r.t.\ $s$.} to decrease exponentially provided that $d_0
>\tilde{\sigma}^2$. Thus, we have
\begin{equation}
\tilde{G}(\calE)\ge\frac{\tilde{d}(\calC)-d-\delta_n}{d_0-d}\ge
\frac{d_1-d-\delta_n}{d_0-d},
\end{equation}
which is positive for $n$ large enough. For example, beyond a certain $n_0$,
it exceeds $\frac{d_1-d}{2(d_0-d)}$, which we take to be
$\alpha(\tilde{\sigma}^2,d,R)$. Now, for a given $\epsilon> 0$, let
$\calT_\epsilon=\{\by:~|\ln\frac{\tilde{G}(\by)}{G(\by)}-nD(\tilde{G}\|G)|\le n\epsilon\}$.
Then, by the weak law of large numbers, $\tilde{G}(\calT_\epsilon)\ge
1-\alpha(\tilde{\sigma}^2,d,R)/2$ for all large $n$. Thus,
\begin{eqnarray}
G(\calE)&\ge&\int_{\calE\cap\calT_\epsilon}G(\by)\mbox{d}\by\\
&=&\int_{\calE\cap\calT_\epsilon}\tilde{G}(\by)e^{-\ln[\tilde{G}(\by)/G(\by)]}\mbox{d}\by\\
&\ge&\tilde{G}(\calE\cap\calT_\epsilon)\cdot\exp\{-n[D(\tilde{G}\|G)+\epsilon]\}\\
&\ge&[\tilde{G}(\calE)-\tilde{G}(\calT_\epsilon^c)]\cdot\exp\{-n[D(\tilde{G}\|G)+\epsilon]\}\\
&\ge&[\alpha(\tilde{\sigma}^2,d,R)-\frac{1}{2}\alpha(\tilde{\sigma}^2,d,R)]
\cdot\exp\{-n[D(\tilde{G}\|G)+\epsilon]\}\\
&=&\frac{1}{2}\alpha(\tilde{\sigma}^2,d,R)
\cdot\exp\{-n[D(\tilde{G}\|G)+\epsilon]\}.
\end{eqnarray}
Since this is true for all $\tilde{\sigma}^2$ with $R_G(d,\tilde{\sigma}^2)>
R$, the tightest bound is obtained by minimizing
\begin{equation}
D(\tilde{G}\|G)=\frac{1}{2}\left[\frac{\tilde{\sigma}^2}{\sigma^2}-
\ln\frac{\tilde{\sigma}^2}{\sigma^2}-1\right],
\end{equation}
in the range $\tilde{\sigma}^2\ge d e^{2R}$, which is attained at
$\tilde{\sigma}^2= d e^{2R}$, yielding the following upper bound on the
exponent:
\begin{equation}
E(R)\le \frac{1}{2}\left[\frac{d e^{2R}}{\sigma^2}-
\ln\frac{d e^{2R}}{\sigma^2}-1\right]=\Ph[R-R_G(d)].
\end{equation}
\qed

\bibliographystyle{is-abbrv}
%\bibliographystyle{plain}

%\bibliography{references}

\end{document}